%% file: main.tex
\documentclass[runningheads]{llncs}
\usepackage{microtype}
\usepackage{graphicx}
\usepackage{booktabs}
\usepackage{colortbl}
\usepackage{pgfplots}
\usepackage{pgfplotstable}
\usepackage{url}
\usepackage{booktabs}
\usepackage{pdfpages}
\usepackage{multirow}
\usepackage[utf8]{inputenc}
\usepackage{authblk}
\usepackage[T1]{fontenc}
\usepackage{amsmath,amssymb}
\usepackage{paralist}
\usepackage{latexsym}
\usepackage{xcolor}
\usepackage{wasysym}
\usepackage{graphicx}
\usepackage{MnSymbol}
\usepackage{stackengine}
\usepackage{mdframed}
\usepackage{comment}
\usepackage{array}

\usepackage{algorithmic}

\usepackage{tikz,color,float,caption,subcaption}
            
\newcommand{\uproman}[1]{\uppercase\expandafter{\romannumeral#1}}

\renewcommand{\figurename}{Fig.}
\renewcommand{\tablename}{Tab.}
\input{commands}

\begin{document}
\title{How Do Practitioners Interpret Conditionals in Requirements?}
%
%
\author{Jannik Fischbach\inst{1} \and
Julian Frattini\inst{2} \and
Daniel Mendez\inst{2,3}\and
Michael Unterkalmsteiner\inst{2} \and
Henning Femmer\inst{1} \and
Andreas Vogelsang\inst{4}}
\authorrunning{J. Fischbach et al.}
%
\institute{Qualicen GmbH, Germany, \email{\{firstname.lastname\}@qualicen.de} \and
Blekinge Institute of Technology, Sweden, \email{\{firstname.lastname\}@bth.se} \and fortiss GmbH, Germany, \email{mendez@fortiss.org} \and University of Cologne, Germany, \email{vogelsang@cs.uni-koeln.de} }
\maketitle              
\begin{abstract}
\textit{Context}: Conditional statements like ``If A and B then C'' are core elements for describing software requirements. However, there are many ways to express such conditionals in natural language and also many ways how they can be interpreted. We hypothesize that conditional statements in requirements are a source of ambiguity, potentially affecting downstream activities such as test case generation negatively. 
\textit{Objective}: Our goal is to understand how specific conditionals are interpreted by readers who work with requirements.
\textit{Method}: We conduct a descriptive survey with 104 RE practitioners and ask how they interpret 12 different conditional clauses. We map their interpretations to logical formulas written in Propositional (Temporal) Logic and discuss the implications. 
\textit{Results}: The conditionals in our tested requirements were interpreted ambiguously. We found that practitioners disagree on whether an antecedent is only \textit{sufficient} or also \textit{necessary} for the consequent. Interestingly, the disagreement persists even when the system behavior is known to the practitioners. We also found that certain cue phrases are associated with specific interpretations. 
\textit{Conclusion}: Conditionals in requirements are a source of ambiguity and there is not just one way to interpret them formally. This affects any analysis that builds upon formalized requirements (e.g., inconsistency checking, test-case generation). Our results may also influence guidelines for writing requirements. 

\keywords{Logical Interpretation  \and Requirements Engineering \and Descriptive Survey \and Formalization.}
\end{abstract}
\section{Introduction}
\textbf{Context} Functional requirements often describe external system behavior by relating events to each other, e.g. ``If the system detects an error ($e_1$), an error message shall be shown ($e_2$)'' (REQ~1). Such conditional statements are prevalent in both traditional requirement documents~\cite{fischbachREFSQ} and agile requirement artifacts~\cite{fischbachICST} alike. The interpretation of the semantics of conditionals affects all activities carried out on the basis of documented requirements such as manual reviews, implementation, or test case generations. Even more, a correct interpretation is absolutely essential for all automatic analyses of requirements that consider the semantics of sentences; for instance, automatic quality analysis like smell detection~\cite{femmer2017rapid}, test case derivation~\cite{fischbachICST,frattini2020}, and dependency detection~\cite{fischbachRE}. In consequence, conditionals should always be associated with a formal meaning to automatically process them. However, determining a suitable formal interpretation is challenging because conditional statements in natural language tend to be ambiguous. Literally, REQ~1 from above may be interpreted as a logical implication ($e_1 \Rightarrow e_2$), in which $e_1$ is a \textit{sufficient} precondition for $e_2$. However, it is equally reasonable to assume that the error message shall not be shown if the error has not been detected (i.e., $e_1$ is a \textit{sufficient} and also \textit{necessary} condition for $e_2$).
Furthermore, it is reasonable to assume that $e_1$ must occur \textit{before} $e_2$. Both assumptions are not covered by an implication as it neglects temporal ordering. In contrast, the assumptions need to be expressed by temporal logic (e.g., LTL~\cite{leadsTo94}). Existing guidelines for expressing requirements have different ways of interpreting conditionals; for instance, Mavin et al.~\cite{mavin09} propose to interpret conditionals as a logical equivalence ($e_1 \Leftrightarrow e_2$) to avoid ambiguity. We argue that the ``correct'' way of interpretation should not just be defined by the authors of a method, but rather from the view of practitioners. This requires an understanding how these interpret such conditionals. Otherwise, we choose a formalization that does not reflect how practitioners interpret conditional sentences, rendering downstream activities error-prone. That is, we would likely derive incomplete test cases or interpret dependencies between the requirements incorrectly.

\textbf{Problem} We lack knowledge on how practitioners interpret conditional statements in requirements and how these interpretations should be formalized accordingly. Moreover, we are not aware of the factors that influence the logical interpretation of conditional clauses in requirements.

\textbf{Contribution} In this paper, we report on a survey we conducted with 104~RE practitioners and determine how they interpret conditional clauses in requirements. The goal of our research is to provide empirical evidence for whether a common formal interpretation of conditionals in requirements exists. Key insights include, but are not limited to:

\begin{compactenum}
  \item Conditionals in requirements are ambiguous. Practitioners disagreed on whether an antecedent is only \textit{sufficient} or also \textit{necessary} for a consequent. 
  \item We observed a statistically significant relation between the interpretation and certain context factors of practitioners (e.g., experience in RE, the way how a practitioner interacts with requirements, and the presence of domain knowledge). Interestingly, domain knowledge does not promote a consistent interpretation of conditionals.
  \item The choice of certain cue phrase has an impact on the degree of ambiguity (e.g., ``while'' was less ambiguous than ``if'' or ``when'' w.r.t. temporal relationship).
\end{compactenum}
Finally, we disclose all of our data as well as the survey protocol via a replication package at \url{https://doi.org/10.5281/zenodo.5070235}.

\textbf{Related Work} 
Transforming NL requirements into verifiable LTL patterns~\cite{dwyer1999patterns} has received notable attention, as this formalization respects the temporal aspect of requirements and allows for an automatic assessment of requirements quality like ambiguity, consistency, or completeness~\cite{nikora2009automated}. However, most approaches are based on restricted natural language~\cite{schumann2020generation,yan2015formal,ghosh2014automatically} and assume that a unanimously agreed upon formalization of NL requirements exist. We challenge this assumption by considering ambiguity in respect to conditional statements. 
Ambiguity in NL requirements itself has been explored in several studies so far. A general overview of the nature of ambiguity and its impact on the development process is provided by Gervasi and Zowghi~\cite{gervasi10}. De Bruijn et al.~\cite{de2010ambiguity} investigates the effects of ambiguity on project success or failure. Berry and Kamsties~\cite{berry05} show that indefinite quantifiers can lead to misunderstandings. Winter et al.~\cite{winter20} show that negative phrasing of quantifiers is more ambiguous than affirmative phrasing. Femmer et al.~\cite{femmer14} reveals that the use of passive voice leads to ambiguity in requirements. To the best of our knowledge, however, we are the first to study ambiguity induced by conditionals in requirements

\section{Fundamentals}\label{fundamentals}
To determine how to appropriately formalize interpretations by RE practitioners, we first need to understand how conditionals can be specified logically. We investigate the logical interpretations with respect to two dimensions: Necessity and Temporality. In this section, we demarcate both dimensions, and introduce suitable formal languages to the extent necessary in context of this paper. 

\textbf{Necessity} A conditional statement consists of two parts: the antecedent (in case of REQ~1: $e_1$) and the consequent ($e_2$). The relationship between an antecedent and consequent can be interpreted logically in two different ways. First, by means of an implication as $e_1 \Rightarrow e_2$, in which $e_1$ is a \textit{sufficient} condition for $e_2$. Interpreting REQ~1 as an implication requires the system to display an error message if $e_1$ is true. However, it is not specified what the system should do if $e_1$ is false. The implication allows both the occurrence of $e_2$ and its absence if $e_1$ is false. In contrast, the relationship of antecedent and consequent can also be understood as a logical equivalence, where $e_1$ is both a \textit{sufficient} and \textit{necessary} condition for $e_2$. Interpreting REQ~1 as an equivalence requires the system to display an error message \textit{if and only if} it detects an error. Consequently, if $e_1$ is false, then $e_2$ should also be false. The interpretation of conditionals as an implication or equivalence significantly influences further development activities. For example, a test designer who interprets conditionals rather as implication than equivalence might only add positive test cases to a test suite. This may lead to a misalignment of tests and requirements in case the business analyst actually intended to express an equivalence.

\textbf{Temporality} 
The temporal relation between an antecedent and consequent can be interpreted in three different ways: (1) the consequent occurs simultaneous with the antecedent, (2) the consequent occurs immediately after the antecedent, and (3) the consequent occurs at some indefinite point after the antecedent. Propositional logic does not consider temporal ordering of events and is therefore not expressive enough to model temporal relationships. In contrast, we require linear temporal logic (LTL), which considers temporal ordering by defining the behavior $\sigma$ of a system as an infinite sequence of states $\langle s_0, \dots \rangle$, where $s_n$ is a state of the system at ``time'' $n$~\cite{leadsTo94}. Accordingly, requirements are understood as constraints on $\sigma$. The desired system behavior is defined as an LTL formula $F$, where next to the usual PL operators also temporal operators like $\square$ (\textit{always}), $\lozenge$ (\textit{eventually}), and $\ocircle$ (\textit{next state}) are used. Since we will use these temporal operators in the course of the paper, we will present them here in more detail. To understand the LTL formulas, we assign a semantic meaning $\lsem F \rsem$ to each syntactic object $F$. Formally, $\lsem F \rsem$ is a boolean-valued function on $\sigma$. According to Lamport~\cite{leadsTo94}, $\sigma \lsem F \rsem$ denotes the boolean value that formula $F$ assigns to behavior $\sigma$, and that $\sigma$ satisfies $F$ if and only if $\sigma \lsem F \rsem$ equals true (i.e., the system satisfies requirement $F$). We define $\lsem \square F \rsem$, $\lsem \lozenge F \rsem$ and $\lsem \ocircle F \rsem$ in terms of $\lsem F \rsem$ (see equations below). The expression $\langle s_0,\dots \rangle \lsem F \rsem$ asserts that $F$ is true at ``time'' 0 of the behavior, while $\langle s_n,\dots \rangle \lsem F \rsem$ asserts that $F$ is true at ``time'' $n$.
\begin{equation}\label{eq:always}
\forall n \in \mathbb{N} : \langle s_n,\dots \rangle \lsem \square F \rsem
\Rightarrow \forall m \in \mathbb{N}, m \geq n, \langle s_m,\dots \rangle \lsem F \rsem
\end{equation}
\begin{equation}\label{eq:eventually}
\forall n \in \mathbb{N} : \langle s_n,\dots \rangle \lsem \lozenge F \rsem
\Rightarrow \exists m \in \mathbb{N}, m > n, \langle s_m,\dots \rangle \lsem F \rsem
\end{equation}
\begin{equation}\label{eq:nextState}
\forall n \in \mathbb{N} : \langle s_n,\dots \rangle \lsem \ocircle F \rsem
\Rightarrow \langle s_{n+1},\dots \rangle \lsem F \rsem
\end{equation}

Equation~\ref{eq:always} asserts that $F$ is true in all states of behavior $\sigma$. More specifically, $\square F$ asserts that $F$ is \textit{always} true. The temporal operator $\lozenge$ can be interpreted as ``it is not the case that $F$ is always false''~\cite{leadsTo94}. According to equation~\ref{eq:eventually}, a behavior $\sigma$ satisfies $\lozenge F$ if and only if $F$ is true at some state of $\sigma$. In other words, $\lozenge F$ asserts that $F$ is \textit{eventually} true. According to equation 3, $\ocircle F$ asserts that $F$ is true at the \textit{next state} of behavior $\sigma$. In contrast to $\lozenge F$, $\ocircle F$ requires that this state is not an arbitrary state of behavior $\sigma$, but rather the direct successor of state $n$. In conclusion, LTL can be used to incorporate temporal ordering into an implication ($F \Rightarrow G$) in three ways:

\begin{compactenum}
  \item $G$ occurs simultaneous with $F$: \\ $\square ( F \Rightarrow G)$, which can be interpreted as ``any time F is true, G is also true''. 
  \item $G$ occurs immediately after $F$: \\ $\square ( F \Rightarrow \ocircle G)$, which can be  interpreted as ``G occurs after F terminated''.
  \item $G$ occurs at some indefinite point after $F$: \\ $\square ( F \Rightarrow \lozenge G)$, which can be  interpreted as ``any time F is true, G is also true or at a later state''.
\end{compactenum}

\begin{figure*}
\centering
    \fbox{\includegraphics[width=\textwidth]{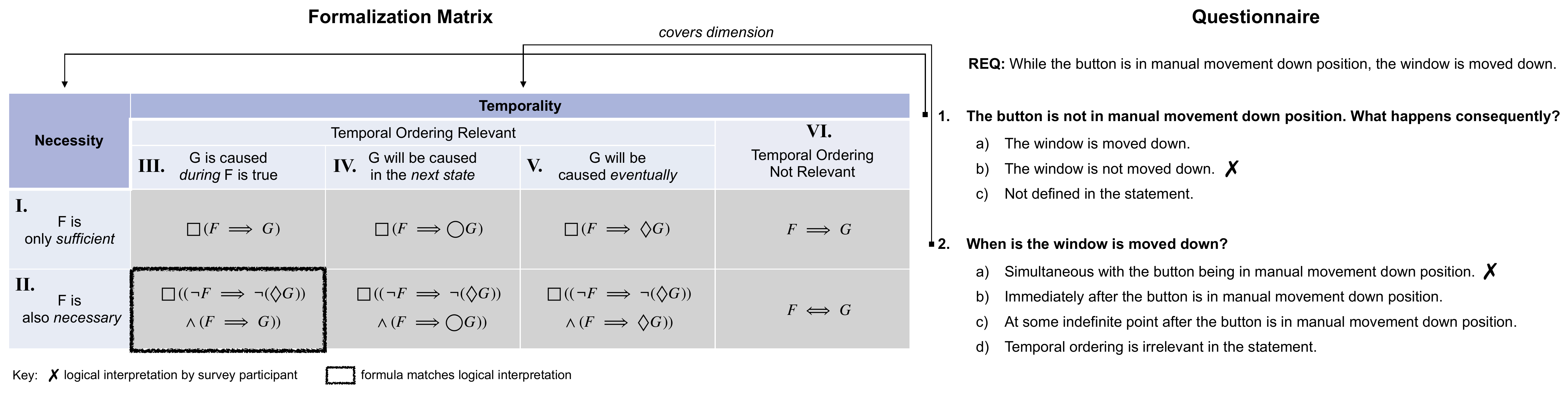}}
    \caption{Mapping between Questionnaire (right) and Formalization Matrix (left).}
    \label{fig:matrix_questions}
     \vspace{-.5cm}
\end{figure*}

\textbf{Formalization Matrix}
To distinguish the logical interpretations and their formalization, we constructed a formalization matrix (see Fig.~\ref{fig:matrix_questions}). It defines a conditional statement of $F$ and $G$ along the two dimensions (Necessity, and Temporality), each divided on a nominal scale (see Tab.~\ref{tab:variables}). Each 2-tuple of characteristics can be mapped to an entry in the formalization matrix. For example, the LTL formula $\square(F \Rightarrow \ocircle G)$ formalizes a conditional statement, in which $F$ is only \textit{sufficient} and $G$ occurs in the \textit{next state}. Conditional statements that define $F$ as both \textit{sufficient} and \textit{necessary} must be formalized with a further LTL formula: $\square ( \neg F \Rightarrow \neg ( \lozenge G))$. This formula can be literally interpreted as ``If $F$ does not occur, then $G$ does not occur either (not even \textit{eventually})''. 
\section{Study Design}
To understand how practitioners interpret conditionals in requirements and how their interpretations should be formalized accordingly, we conducted a survey following the guidelines by Ciolkowski et al.~\cite{Ciolkowski2003}.

\subsection{Survey Definition}

We aim to understand and (logically) formalize the interpretation of conditionals in requirements by RE practitioners in software development projects.
The expected outcome of our survey is a better understanding of how practitioners logically interpret conditional clauses in requirements and which of the elements in our formalization matrix match their logical interpretations (see Fig.~\ref{fig:matrix_questions}). We derived three research questions (RQ) from our survey goal.

\begin{compactitem}
  \item \textbf{RQ1:} How do practitioners logically interpret conditional clauses in requirements?
  \item \textbf{RQ2:} Which factors influence the logical interpretation of conditional clauses in requirements?
  \item \textbf{RQ3:} Which (if any) cue phrases promote (un)ambiguous interpretation?
\end{compactitem}

RQ1 investigates how conditionals are interpreted by practitioners and how their interpretations should be formalized accordingly. RQ2 studies whether the logical interpretation of practitioners depends on certain factors. We focus on: 1) the role of the participant (e.g., writing requirements vs.\ reading and implementing requirements) and 2) the domain context of the requirement (i.e., does the requirement describe system behavior from a domain that is familiar to the participant, or does the requirement originate from an unknown domain?). RQ3 aims at the formulation of conditionals: Conditional clauses can be expressed by using different cue phrases (e.g., ``if'', ``when''). We hypothesize that cue phrases impact the logical interpretation of practitioners. With RQ3, we want to identify cue phrases for which the interpretations are almost consistent, and cue phrases which are ambiguous. This insight enables us to derive best practices on writing conditionals in requirements specifications.

\subsection{Survey Design}

\textbf{Target Population and Sampling}
The selection of the survey participants was driven by a purposeful sampling strategy~\cite{baltes2020sampling} along the following criteria: a) they elicit, maintain, implement, or verify requirements, and b) they work in industry and not exclusively in academia.
Each author prepared a list of potential participants using their personal or second-degree contacts (convenience sampling~\cite{wohlin12}). From this list, the research team jointly selected suitable participants based on their adequacy for the study. To increase the sample size further, we asked each participant for other relevant contacts after the survey (snowball sampling). Our survey was started by 168~participants of which 104~completed the survey. All figures in this paper refer to the 104 participants that completed the survey. The majority of participants were non-native English speakers (94.2\%). We received responses mainly from practitioners working in Germany (94.2\%). The remaining 5.8\% of survey completions originate from Croatia, Austria, Japan, Switzerland, United States, and China. The experience of the participants in RE and RE-related fields is equally distributed: 18.2\% have less than 1 year experience, 26\% between 1 and 3 years, 25\% between 4 and 10 years, and 30.8\% more than 10 years. The participants work for companies operating in 22 different domains. The majority of our participants is employed in the automotive (21\%) and insurance/reinsurance (10.1\%) industry. Over the past three years, our participants have worked in 18 different roles. Most frequently, they had roles as developers, project managers, requirements engineers/business analysts, or testers. 77.9\% of the survey participants elicit requirements as part of their job. 59.6\% verify whether requirements are met by a system. 46.2\% read requirements and implement them. 45.2\% maintain the quality of requirements.

\textbf{Study Objects}
To conduct the survey and answer the RQs, we used three data sets (DS), each from a different domain. DS1 contains conditionals from a requirements document describing the behavior of an automatic door in the automotive domain. We argue that all participants have an understanding of how an automatic car door is expected to work, so that all participants should have the required domain knowledge. DS2 contains conditionals from aerospace systems. 
We hypothesize that no or only few participants have deeper knowledge in this domain, making DS2 well suited for an analysis of the impact of domain knowledge on logical interpretations. DS3 contains abstract conditionals (e.g., If event A and event B, then event C). Thus, they are free from any domain-induced interpretation bias. To address RQ~3, we focused on four cue phrases in the conditionals: ``if'', ``while'', ``after'', and ``when''. To avoid researcher bias, we created the datasets extracting conditionals randomly from existing requirement documents used in practice. 
The conditionals in DS1 are taken from a requirements document written by Mercedes-Benz Passenger Car Development.\footnote{Thanks to Frank Houdek for sharing the document at NLP4RE'19~\cite{nlp4re}: \url{https://nlp4re.github.io/2019/uploads/demo-spec-automatic-door.pdf}} 
The conditionals contained in DS2 originate from three requirements documents published by NASA and one by ESA.\footnote{We retrieved these documents from the data set published by Fischbach~et~al.~\cite{fischbachRE}. We are referring to the documents: REQ-DOC-22, REQ-DOC-26, REQ-DOC-27 and REQ-DOC-30.} 
The conditionals in DS3 are syntactically identical to the conditionals in DS1, except that we replace the names of the events with abstract names. DS1--3 contain four conditionals each, resulting in a total of 12 study objects. Each cue phrase occurs exactly once in each DS.

\textbf{Questionnaire Design}
We chose an online questionnaire as our data collection instrument to gather quantitative data on our research questions. For the design, we followed the guidelines of Dillman~et~al.~\cite{Dillman14} to reduce common mistakes when setting up a questionnaire. 
Since our research goal is of descriptive nature, most questions are closed-ended. We designed three types of questions (Q) addressing the two dimensions and prepared a distinct set of responses (R), among which the participants can choose. Each of these responses can be mapped to a characteristic in the formalization matrix and thus allows us to determine which characteristic the practitioners interpret as being reflected by a conditional (see Fig.~\ref{fig:matrix_questions}). 
We build the questionnaire for each study object (e.g., If $F$ then $G$) according to a pre-defined template (see Fig.~\ref{fig:questions}). The template is structured as follows: The first question (\textbf{Q1}) investigates the dimension of Necessity: if event $G$ cannot occur without event $F$, then $F$ is not only \textit{sufficient}, but also necessary for $G$. We add ``nevertheless'' as a third response option (see R.1.1 in Fig.~\ref{fig:questions}) to perform a sanity check on the answers of the respondents. We argue that interpreting that the consequent should occur although the antecedent does not occur indicates that the sentence has not been read carefully. The second question (\textbf{Q2}) covers the temporal ordering of the events. In this context, we explicitly ask for the three temporal relations \textit{eventually}, \textit{always} and \textit{next state} described in Section~\ref{fundamentals}. Should a participant perceive temporal ordering as irrelevant for the interpretation of a certain conditional, we can conclude that PL is sufficient for its formalization. We ask \textbf{Q1--2} for each of the 12 study objects, resulting in a total of 24 questions. To get an overview of the background of our respondents, we also integrated five demographic questions. In total, our final questionnaire consists of 29 questions and can be also found in our replication package.
\begin{figure}
\begin{mdframed}
\scriptsize
    \textbf{Q1}: $F$ does not occur. What happens consequently?
    \begin{itemize}
        \item \textbf{R1.1}: $G$ occurs nevertheless. (sanity check)
        \item \textbf{R1.2}: $G$ does not occur. ($\rightarrow$ \uproman{2})
        \item \textbf{R1.3}: Not defined in the statement. ($\rightarrow$ \uproman{1})
    \end{itemize}
     \textbf{Q2}: When does $G$ occur?
    \begin{itemize}
        \item \textbf{R2.1}: Simultaneously with $F$. ($\rightarrow$ \uproman{3})
        \item \textbf{R2.2}: Immediately after $F$. ($\rightarrow$ \uproman{4})
        \item \textbf{R2.3}: At some indefinite point after $F$. ($\rightarrow$ \uproman{5})
        \item \textbf{R2.3}: Temporal ordering is irrelevant in the statement. ($\rightarrow$ \uproman{6})
    \end{itemize}
\end{mdframed}
\vspace{-.5cm}
\caption{Questionnaire template. The note after each answer option (e.g., $\rightarrow$ \uproman{4}) indicates the matching characteristic in the formalization matrix (see Fig.~\ref{fig:matrix_questions}). If a participant selects R1.2, for example, she implicitly interprets $F$ as \textit{necessary} for $G$. The notes were not included in the questionnaire.}
\label{fig:questions}
 \vspace{-.5cm}
\end{figure}
    
\subsection{Survey Implementation and Execution}
We prepared an invitation letter to ask potential participants if they would like to join our survey. We incorporated all of our 29 questions into the survey tool Unipark~\cite{unipark}. To avoid bias in the survey data, we allow Unipark to randomize the order of the non-demographic questions. We opened the survey on Feb 01, 2021 and closed it after 15 days. We approached all eligible contacts from our prepared list either by e-mail or via Linkedin direct message. We also distributed the questionnaire via a mailing list in the RE focus group of the German Informatics Society (GI). As the traffic on our survey website decreased during the first week, we contacted all candidates again on Feb 08.

\subsection{Survey Analysis}
To answer the proposed research questions, we analyzed the gathered quantitative data as follows.

\textbf{Analysis for RQ 1}
We use heatmaps to visualize how the respondents logically interpret the individual study objects (see Fig.~\ref{fig:RQ1heatmaps}). Each cell in the heatmaps corresponds to a single 2-tuple. Based on the heatmaps, we analyse the logical interpretations of the participants and decide which formalization should be chosen for each study subject according to the most frequent 2-tuple.

\textbf{Analysis for RQ 2}
We focus on three factors ($f_n$) and investigate their impact on the logical interpretations of practitioners: 
(1) the experience in RE ($f_1$: $\boldsymbol{\mathsf{Experience}}$), (2) how the practitioners interact with requirements (elicit, maintain, verify,\ldots) in their job ($f_2$: $\boldsymbol{\mathsf{Interaction}}$), and (3) the domain context of the conditional ($f_3$: $\boldsymbol{\mathsf{Domain}}$). 
To answer RQ 2, we examine the impact of $f_1$--$f_3$ on the dimensions described in Section 2. In our survey, we collected the dimensions for each sentence individually, resulting in 12 categorical variables per dimension (e.g., $\boldsymbol{\mathsf{nec_{s1}}}$, $\boldsymbol{\mathsf{nec_{s2}}}$, \dots $\boldsymbol{\mathsf{nec_{s12}}}$). To get an insight across all sentences, we aggregated all 12 categorical variables per dimension to one variable (resulting in $\boldsymbol{\mathsf{Necessity}}$, and $\boldsymbol{\mathsf{Temporality}}$). This allows us to analyze, for example, whether the experience of the respondents has an impact on understanding an antecedent only as \textit{sufficient} for a consequent or as both \textit{sufficient} and \textit{necessary}. In other words, does the perception of $\boldsymbol{\mathsf{Necessity}}$ depend on $\boldsymbol{\mathsf{Experience}}$? As shown in Table~\ref{tab:variables}, all five variables (3x factors and 2x dimensions) are categorical with a maximum of four levels. The majority is nominally scaled, while $\boldsymbol{\mathsf{Experience}}$ follows an ordinal scale. The variable $\boldsymbol{\mathsf{Domain}}$ was not gathered directly from the responses, but implicitly from our selection of the data sets. We thus add $\boldsymbol{\mathsf{Domain}}$ as variable to our data set, using a categorical scale with three levels: domain knowledge is present (in case of DS1), domain knowledge is not present (DS2), and domain knowledge is not necessary (DS3). By introducing this new variable, we are able to investigate the relationship between domain knowledge and logical interpretations. We use the \textit{chi-squared test of independence} ($\chi^2$) to analyze the relationship between all variables. We run the test by using SPSS and test the following hypotheses ($H_n$):

\begin{algorithmic}
    \FOR{$f_n \in$ \{$\boldsymbol{\mathsf{Experience}}$, $\boldsymbol{\mathsf{Interaction}}$, $\boldsymbol{\mathsf{Domain}}$\}}
        \FOR{$v \in$ \{$\boldsymbol{\mathsf{Necessity}}$, $\boldsymbol{\mathsf{Temporality}}$\}}
            \STATE \textbf{H\textsubscript{0}}: The interpretation of $v$ is independent of $f_n$.
            \STATE \textbf{H\textsubscript{1}}: The interpretation of $v$ depends on $f_n$.
        \ENDFOR
    \ENDFOR
\end{algorithmic}

We set the p-value at 0.05 as the threshold to reject the null hypothesis. To test our hypotheses, we need to calculate the contingency tables for each combination of $f_n$ and dimension. The total number of survey answers per dimension is 1,248 (104 survey completions * 12 annotated sentences). Since we allow the respondents to specify multiple ways to interact with requirements (e.g., to both elicit and implement requirements), our survey data contains a multiple dichotomy set for $\boldsymbol{\mathsf{Interaction}}$. In other words, we created a separate variable for each of the selectable interaction ways (four in total for verify, maintain, elicit and implement). Each variable has two possible values (0 or 1), which indicate whether or not the response was selected by the participant. Therefore, we define a multiple response set in SPSS to create the contingency table for $\boldsymbol{\mathsf{Interaction}}$. The $\chi^2$ test allows us to determine if there is enough evidence to conclude an association between two categorical variables. However, it does not indicate the strength of the relationship. To measure the association between our variables, we use Cramer's Phi $\phi$~\cite{Cohen88} in case of two nominally scaled variables and Freeman's theta $\Theta$~\cite{Freeman65} in case of one ordinally scaled and one nominally scaled variable. We calculate $\phi$ by using SPSS and $\Theta$ by using the R implementation ``freemanTheta''. We interpret $\Theta$ according to the taxonomy of Vargha and Delaney~\cite{Vargha}. For the interpretation of $\phi$, we use the taxonomy of Cohen~\cite{Cohen88}. 

\textbf{Analysis for RQ 3}
A conventional way to measure ambiguity is by calculating the inter-rater agreement (e.g., Fleiss Kappa~\cite{fleiss04}). However, inter-rater agreement measures must be used carefully, as they have a number of well known shortcomings~\cite{FEINSTEIN1990}. For example, the magnitude of the agreement values is not meaningful if there is a large gap between the number of annotated units and the number of involved raters. In our case, we examine only three units per cue phrase (i.e., ``if'' is only included in S2, S8 and S10), each of which was annotated by 104 raters. This discrepancy between the number of units and raters leads to a very small magnitude of the agreement values and distorts the impression of agreement. For example, if we calculate Fleiss Kappa regarding the dimension Temporality of sentences that contain the cue phrase ``while'', we obtain a value of 0.053. According to the taxonomy Landis and Koch~\cite{landis77}, this would imply only a slight agreement between the raters. In fact, however, there is a substantial agreement among the raters that ``while'' indicates a simultaneous relationship. This can be demonstrated by the distribution of survey answers across the different Temporality levels (see Fig.~\ref{fig:RQ3histograms}). 
Thus, instead of reporting less meaningful inter-rater agreement measures, we provide histograms visualizing the distribution of ratings on the three investigated dimensions. We create the histograms for each set of study objects containing the same cue phrase. This allows us to analyze which cue phrase produced the highest/lowest agreement for a certain dimension.
\begin{table}
\scriptsize
\centering
\caption{Overview of analyzed variables.}
\label{tab:variables}
\begin{tabular}{llll} 
\toprule
\textbf{Name}                                  & \textbf{Levels}                                                                                                                                                                                                     & \textbf{Type}                                                          & \textbf{Scale}  \\ 
\hline
$\boldsymbol{\mathsf{Experience}}$  & \begin{tabular}[c]{@{}l@{}}\begin{tabular}{@{\labelitemi\hspace{\dimexpr\labelsep+0.5\tabcolsep}}l}less than 1 year\\1--3 years\\4--10 years\\more than 10 years\end{tabular}\end{tabular}                        & \begin{tabular}[c]{@{}l@{}}categorical\\(single select)\end{tabular}   & ordinal         \\ 
\hline
$\boldsymbol{\mathsf{Interaction}}$ & \begin{tabular}[c]{@{}l@{}}\begin{tabular}{@{\labelitemi\hspace{\dimexpr\labelsep+0.5\tabcolsep}}l}elicit\\maintain\\verify\\implement\end{tabular}\end{tabular}                                                           & \begin{tabular}[c]{@{}l@{}}categorical\\(multiple select)\end{tabular} & nominal         \\ 
\hline
$\boldsymbol{\mathsf{Domain}}$                                         & \begin{tabular}[c]{@{}l@{}}\begin{tabular}{@{\labelitemi\hspace{\dimexpr\labelsep+0.5\tabcolsep}}l}domain knowledge present\\domain knowledge not present\\domain knowledge not necessary\end{tabular}\end{tabular} & \begin{tabular}[c]{@{}l@{}}categorical\\(single select)\end{tabular}   & nominal         \\ 
\hline
$\boldsymbol{\mathsf{Necessity}}$                                            & \begin{tabular}[c]{@{}l@{}}\begin{tabular}{@{\labelitemi\hspace{\dimexpr\labelsep+0.5\tabcolsep}}l}nevertheless\\only sufficient\\also necessary\end{tabular}\end{tabular}                                          & \begin{tabular}[c]{@{}l@{}}categorical\\(single select)\end{tabular}   & nominal         \\ 
\hline
$\boldsymbol{\mathsf{Temporality}}$                                           & \begin{tabular}[c]{@{}l@{}}\begin{tabular}{@{\labelitemi\hspace{\dimexpr\labelsep+0.5\tabcolsep}}l}during\\next state\\eventually\\temporal ordering not relevant\end{tabular}\end{tabular}                  & \begin{tabular}[c]{@{}l@{}}categorical\\(single select)\end{tabular}   & nominal         \\ 
\bottomrule
\end{tabular}
\vspace{-.5cm}
\end{table}
\vspace{-.3cm}
\section{Results}

\subsection*{RQ 1: How do practitioners logically interpret conditional clauses in requirements?}\label{resultsRQ1}

\begin{figure*}
    \scriptsize
    \centering
    \resizebox{\textwidth}{!}{\begin{tabular}{c|c|c}
        \textbf{Automotive Domain (DS1)} & \textbf{Aerospace Domain (DS2)} & \textbf{Abstract Domain (DS3)} \\ \hline
    
        \subcaptionbox*{[S1] \cornIItIV \label{fig:RQ1s1}}{\includegraphics[width=0.43\textwidth]{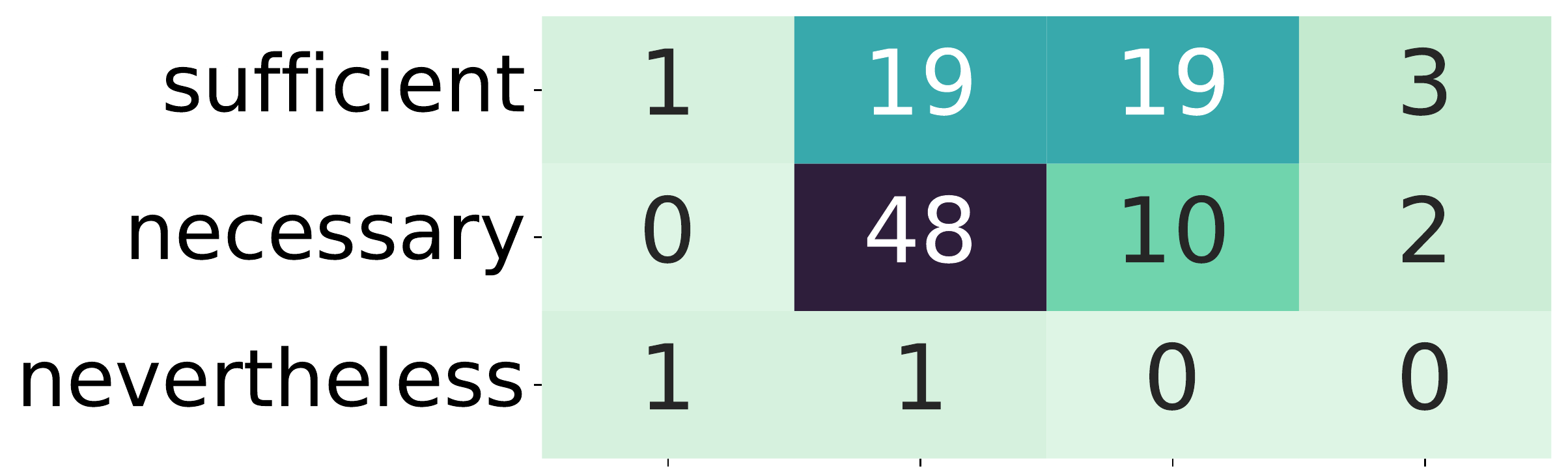}} &
        \subcaptionbox*{[S5] \cornItIII \label{fig:RQ1s5}}{\includegraphics[width=0.285\textwidth]{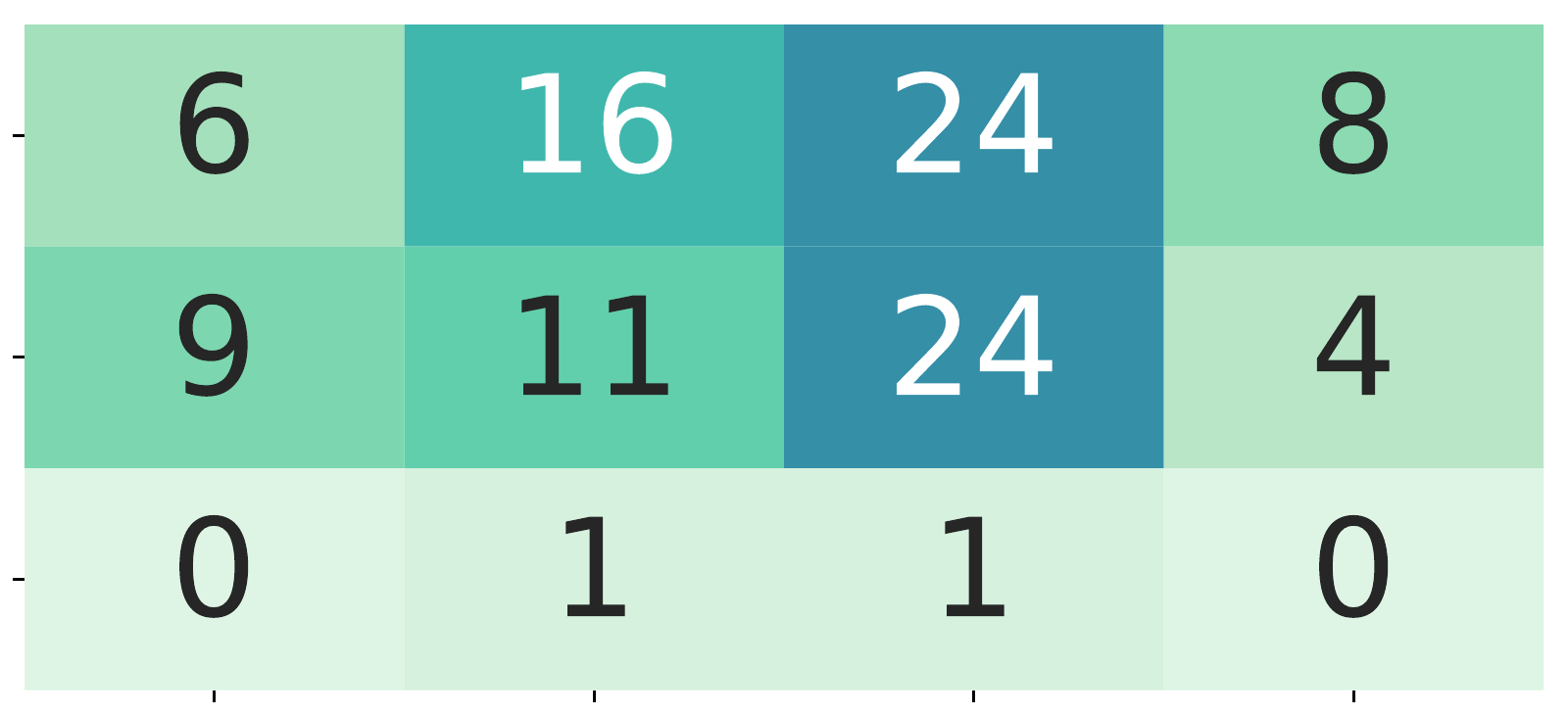}} &
        \subcaptionbox*{[S9] \cornItIII \label{fig:RQ1s9}}{\includegraphics[width=0.285\textwidth]{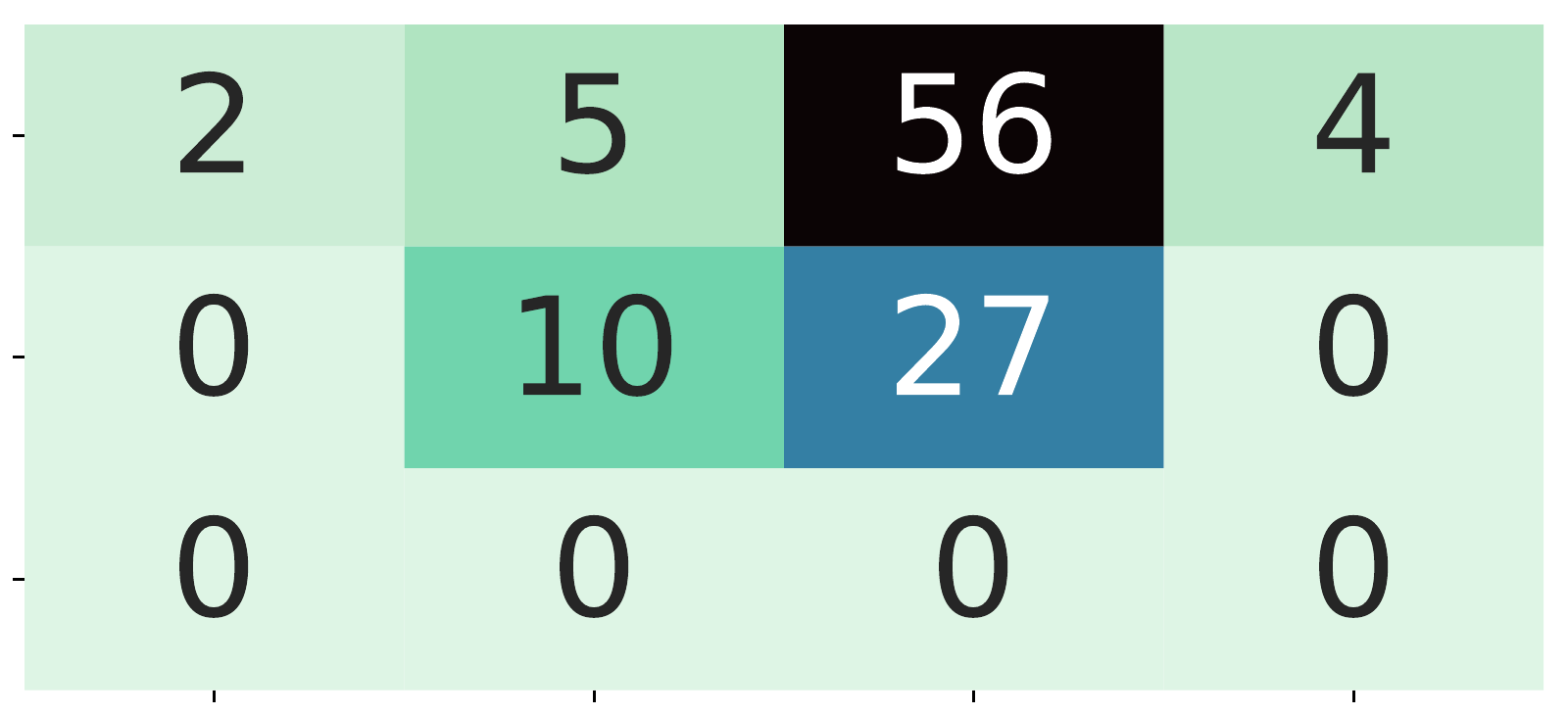}} \\
        
        \subcaptionbox*{[S2] \cornIItVI \label{fig:RQ1s2}}{\includegraphics[width=0.43\textwidth]{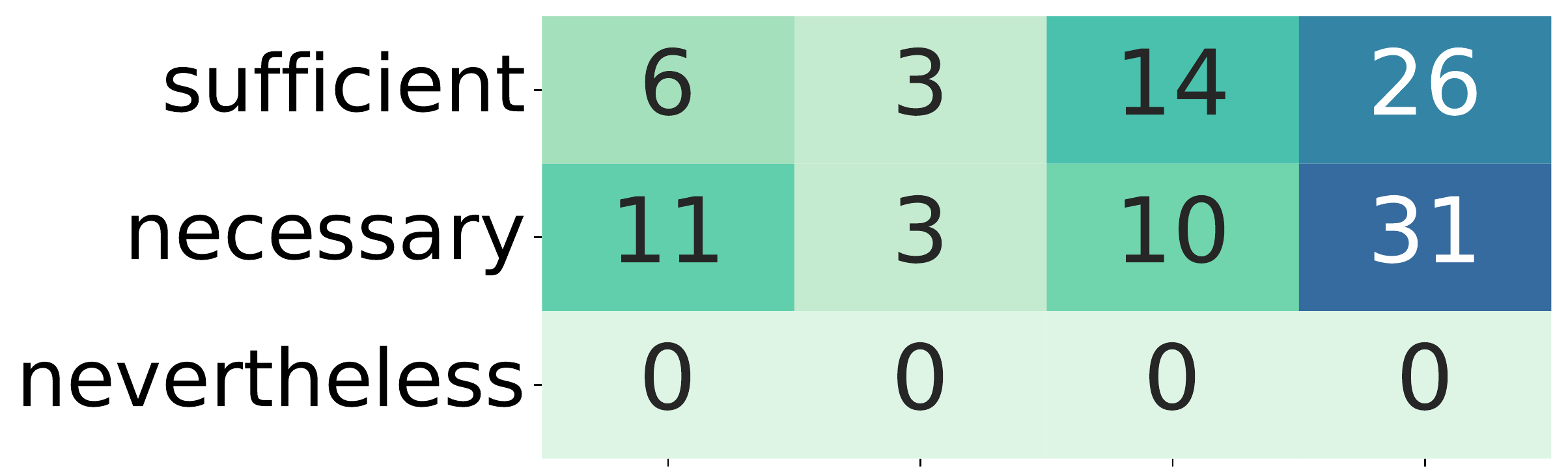}} &
        \subcaptionbox*{[S6] \cornIItV \label{fig:RQ1s6}}{\includegraphics[width=0.285\textwidth]{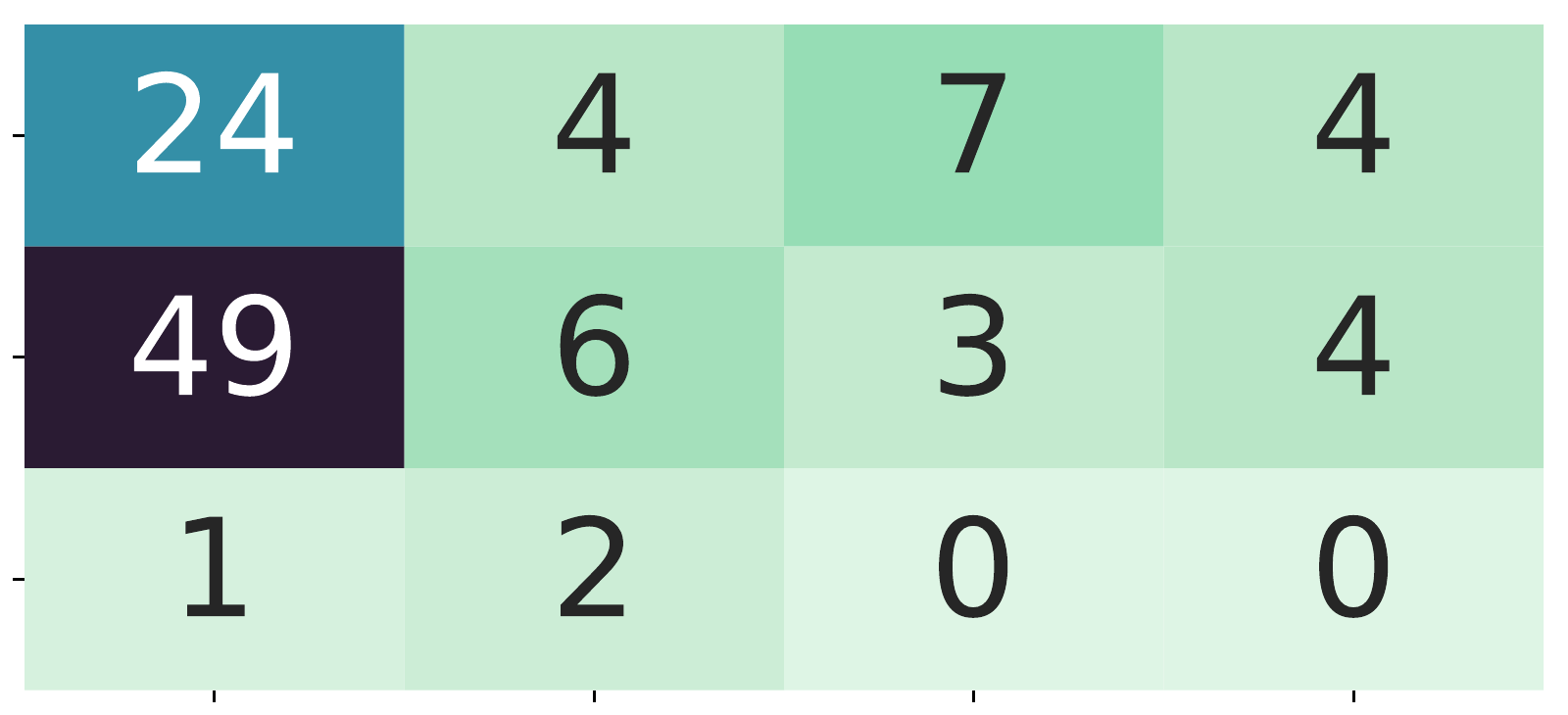}} &
        \subcaptionbox*{[S10] \cornIItV \label{fig:RQ1s10}}{\includegraphics[width=0.285\textwidth]{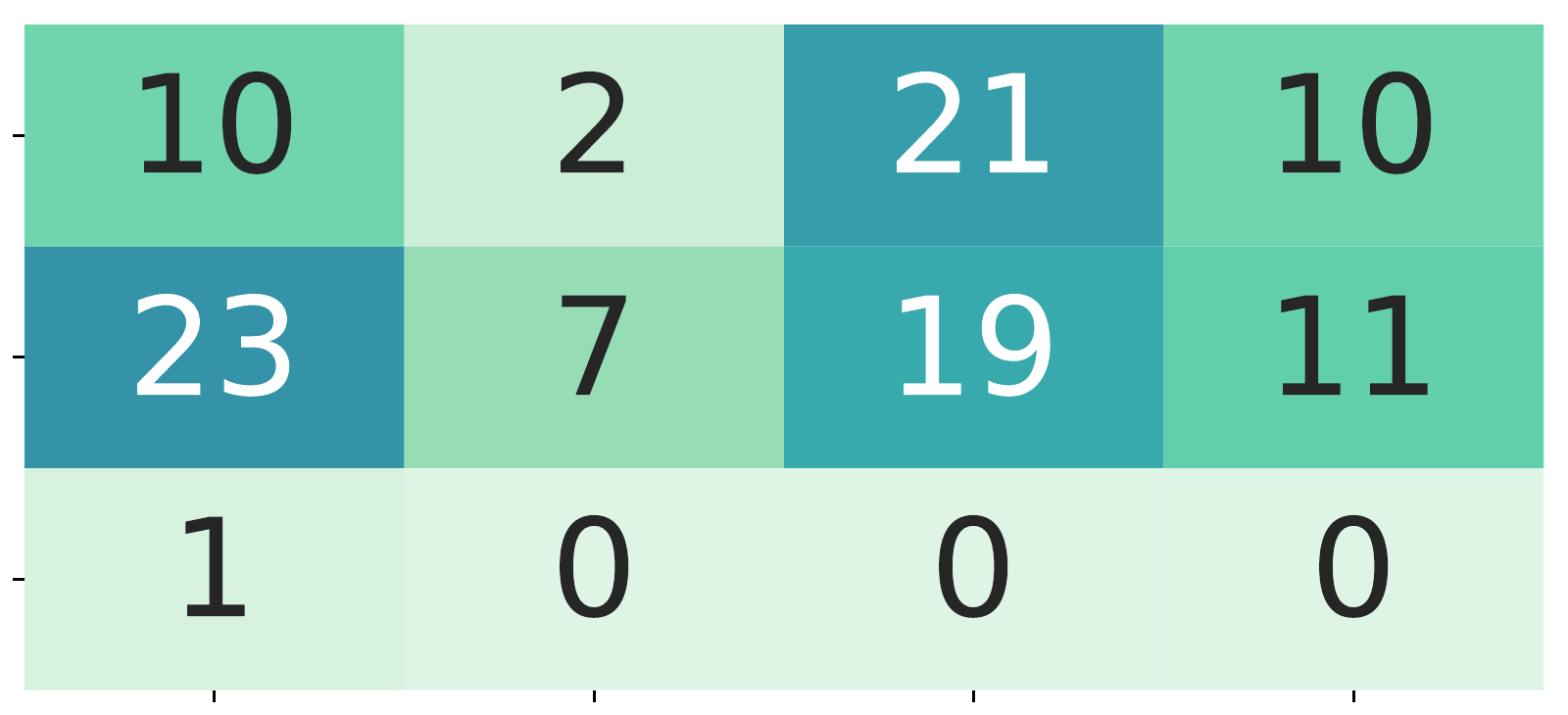}} \\
        
        \subcaptionbox*{[S3] \cornIItIV \label{fig:RQ1s3}}{\includegraphics[width=0.43\textwidth]{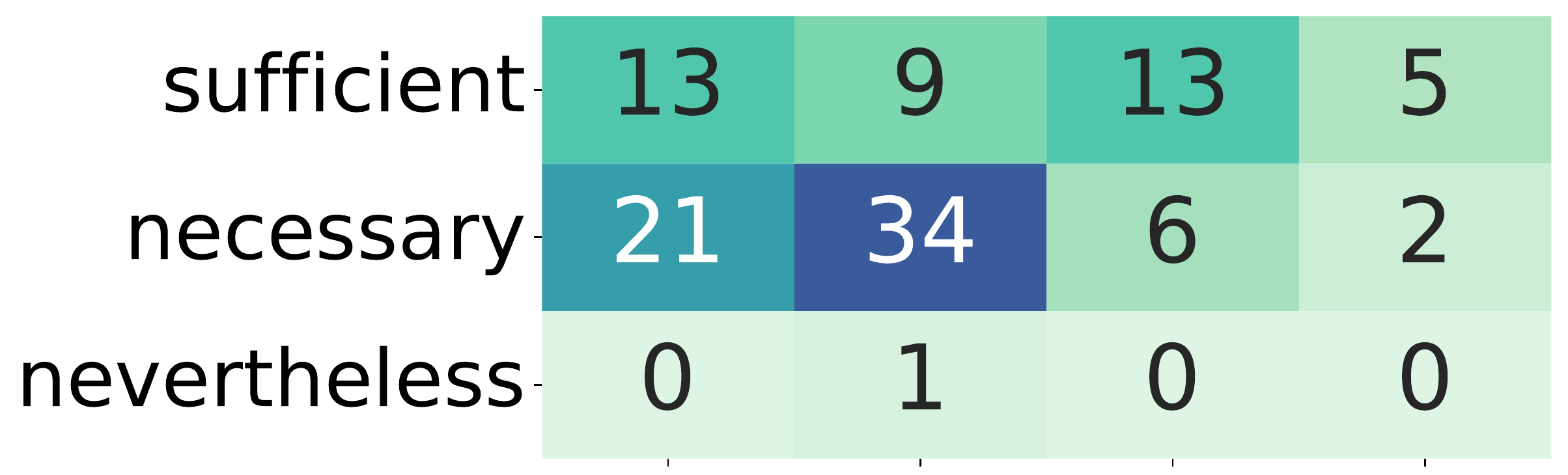}} &
        \subcaptionbox*{[S7] \cornItIII \label{fig:RQ1s7}}{\includegraphics[width=0.285\textwidth]{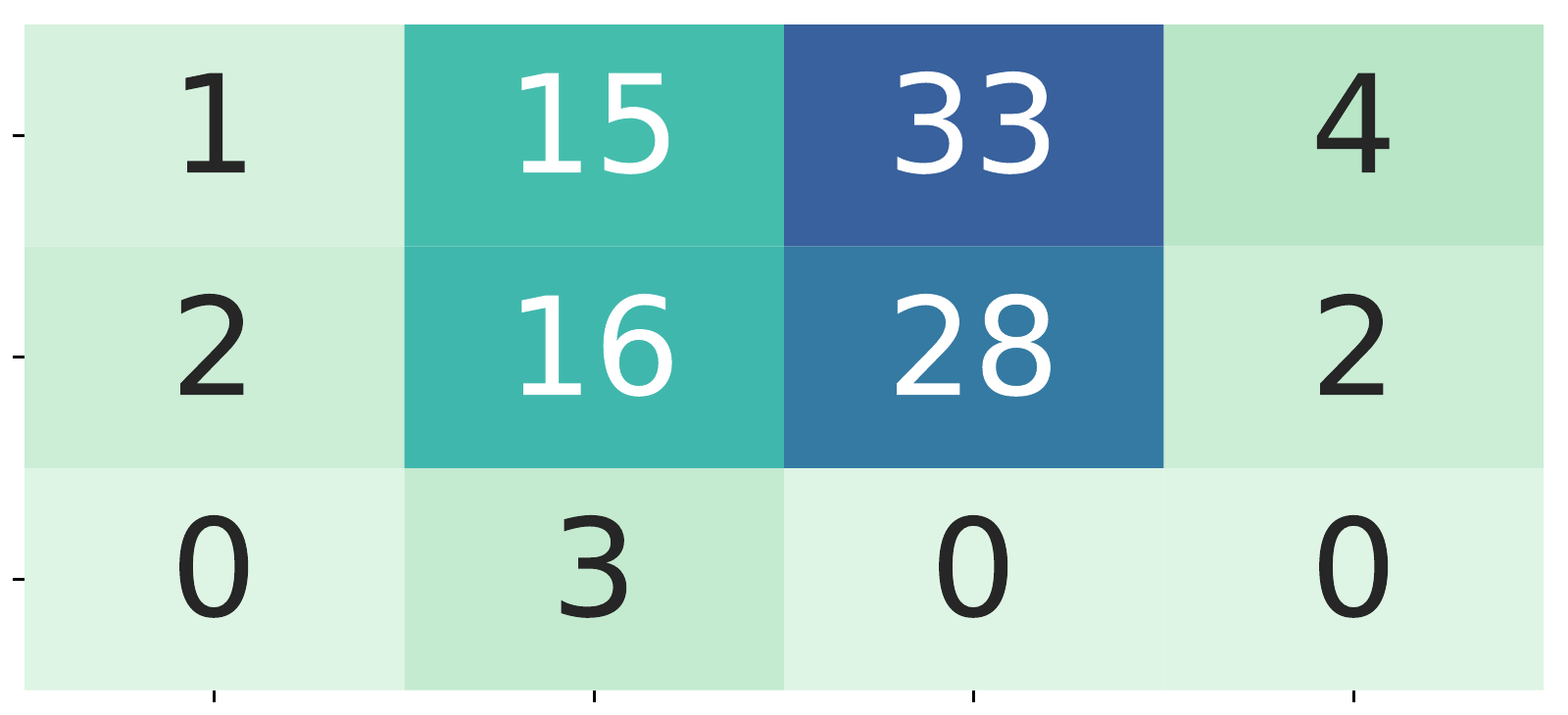}} &
        \subcaptionbox*{[S11] \cornIItV \label{fig:RQ1s11}}{\includegraphics[width=0.285\textwidth]{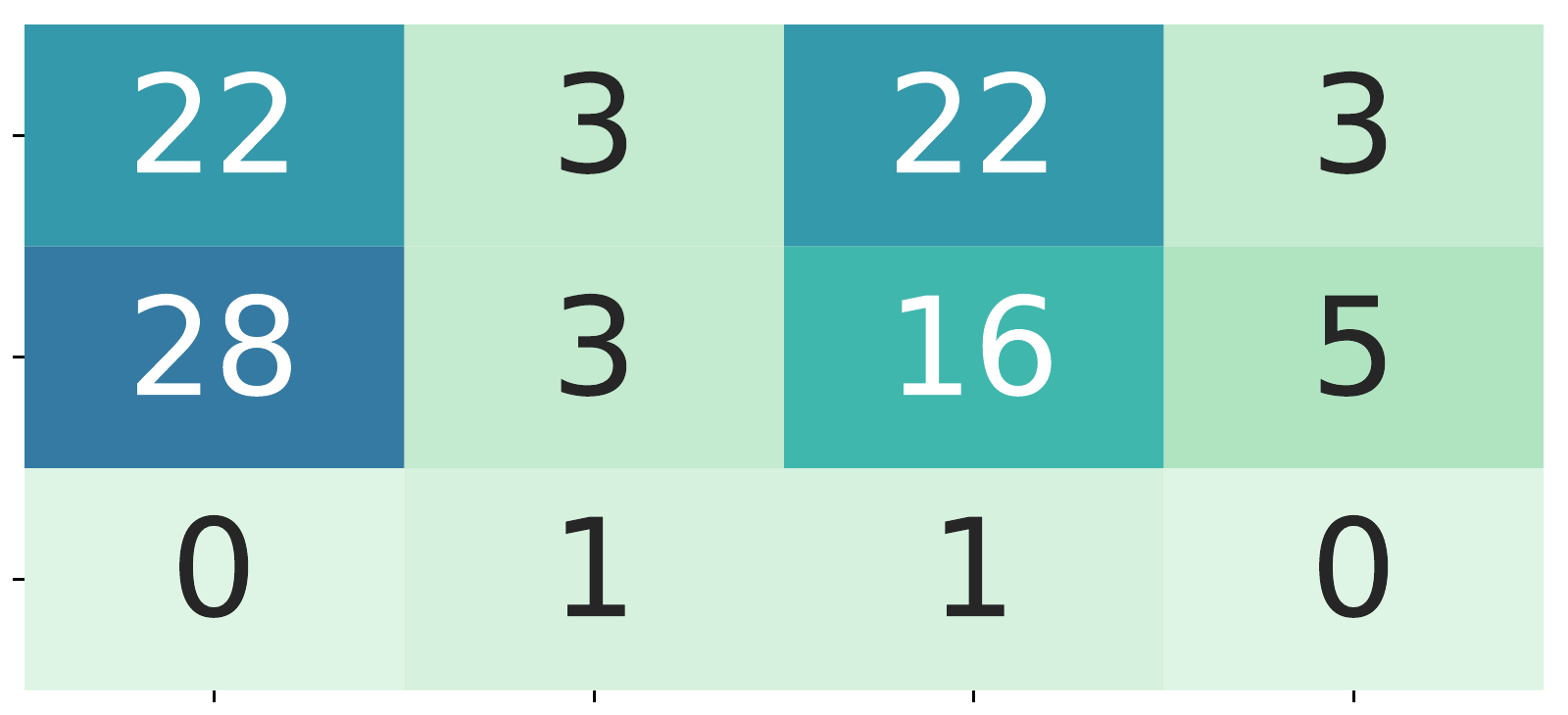}} \\
        
        \subcaptionbox*{[S4] \cornItV \label{fig:RQ1s4}}{\includegraphics[width=0.43\textwidth]{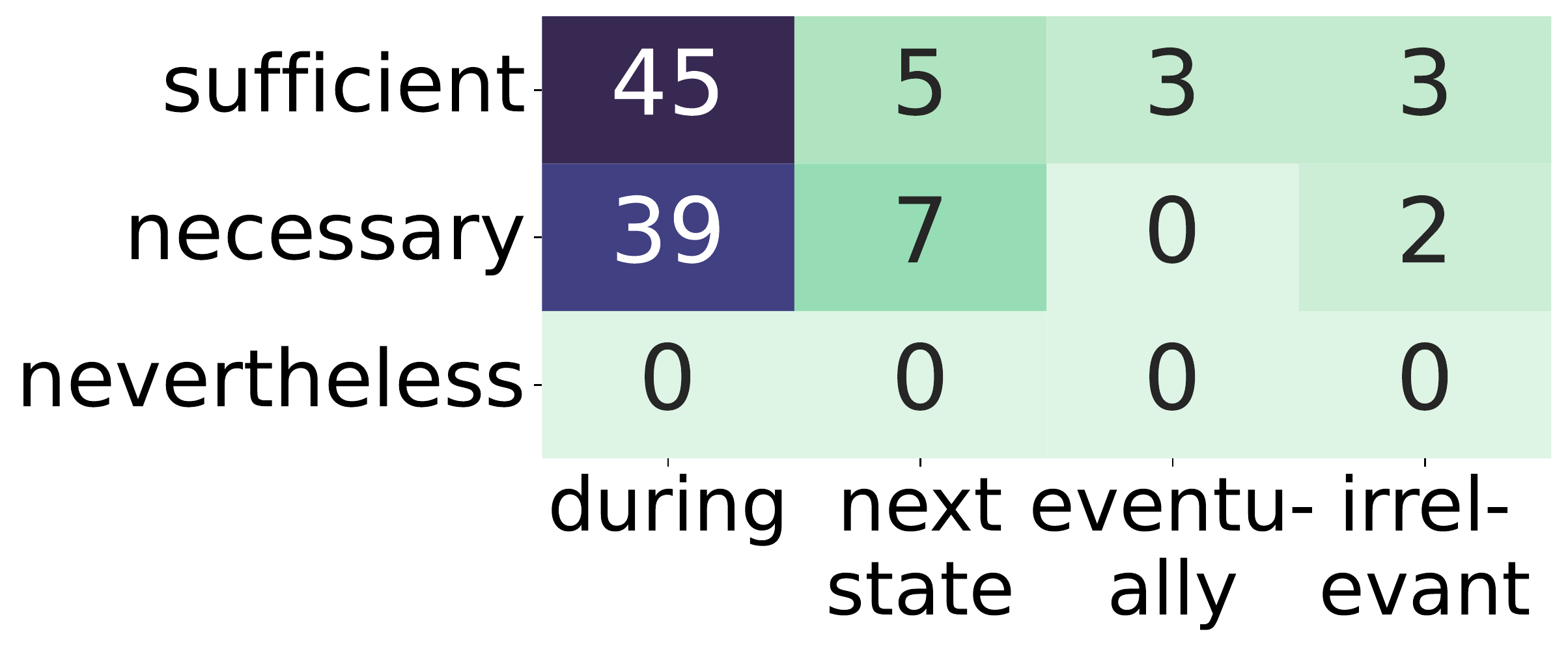}} &
        \subcaptionbox*{[S8] \cornItIII \label{fig:RQ1s8}}{\includegraphics[width=0.285\textwidth]{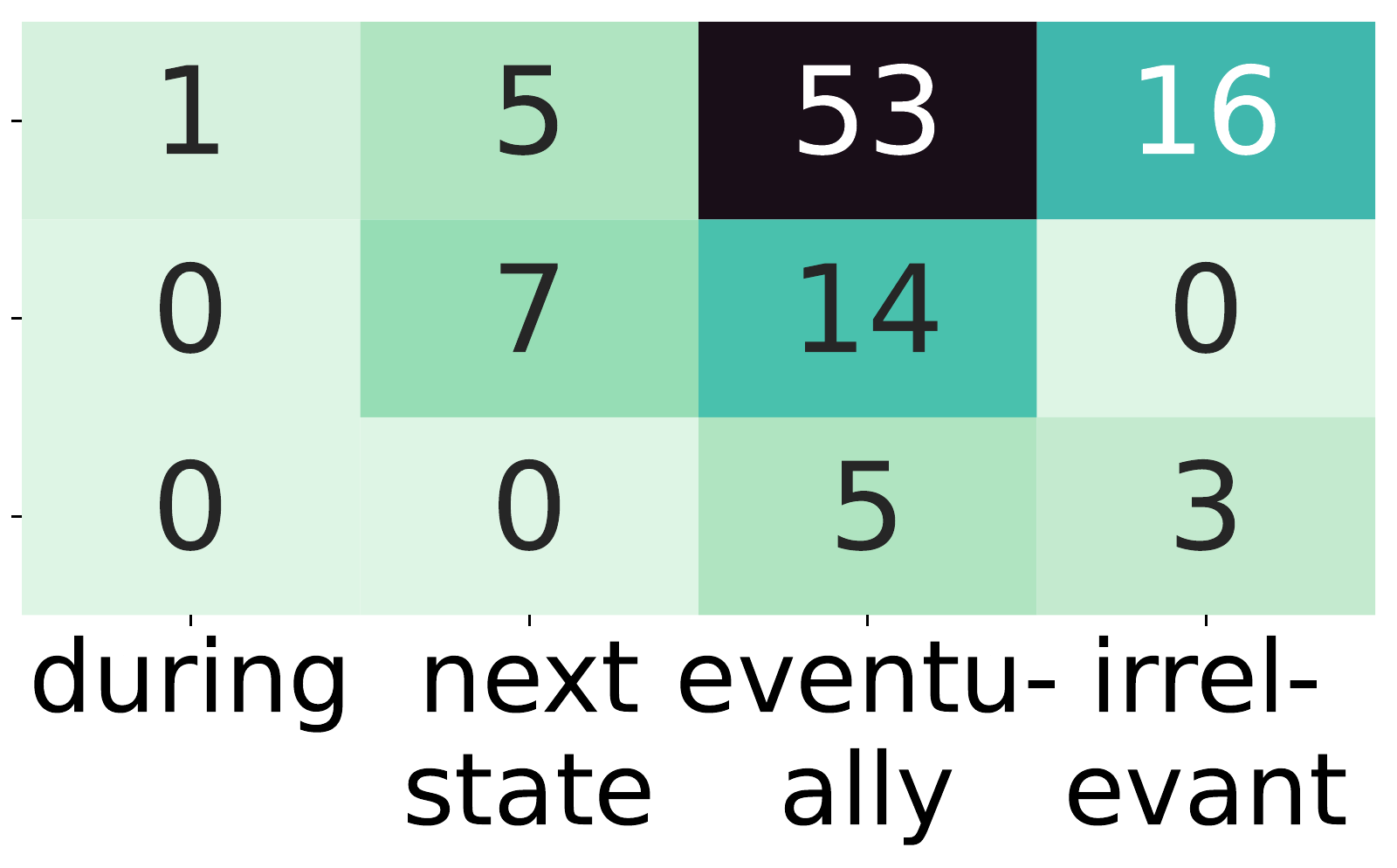}} &
        \subcaptionbox*{[S12] \cornItV \label{fig:RQ1s12}}{\includegraphics[width=0.285\textwidth]{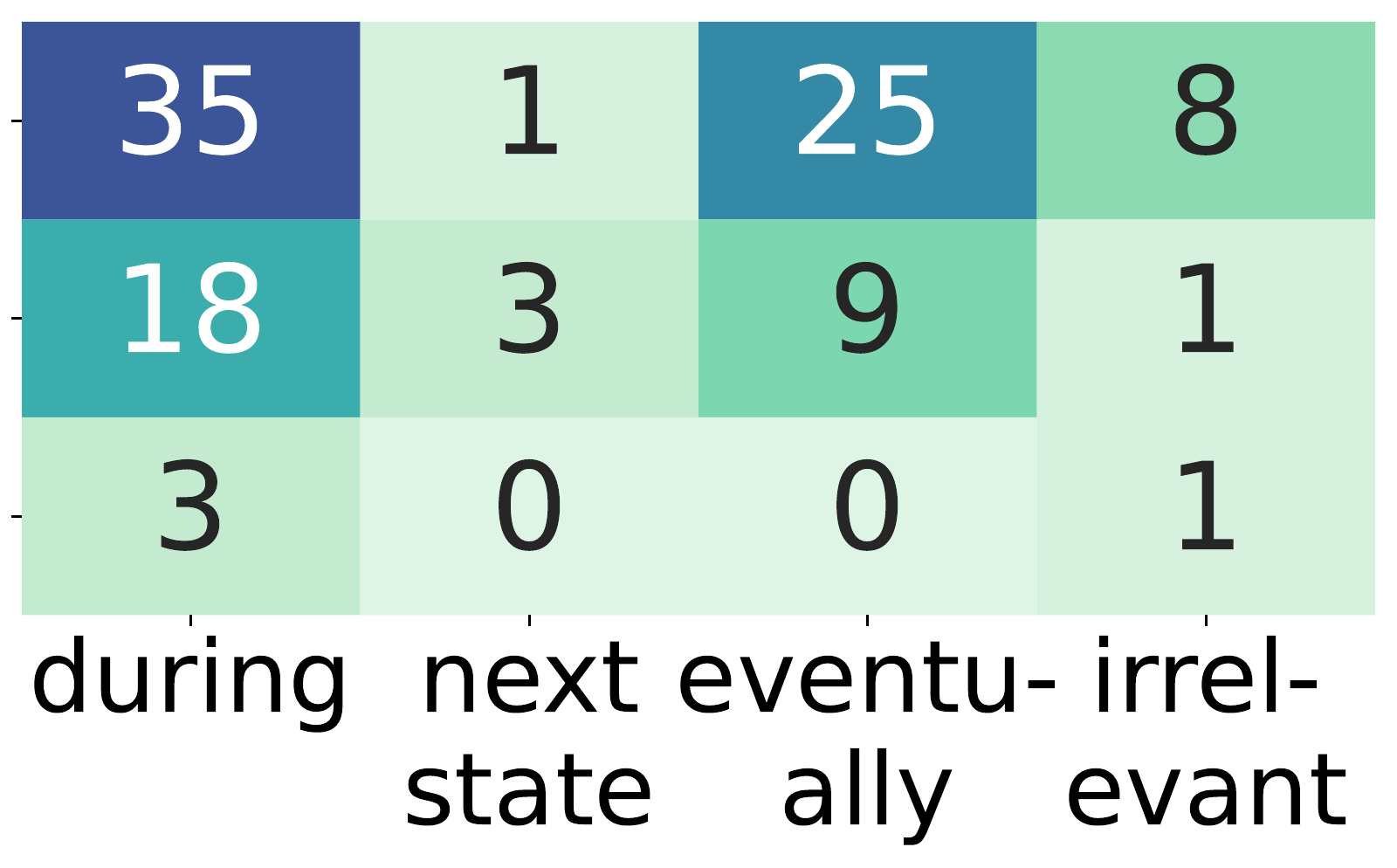}} \\
    \end{tabular}}
    \includegraphics[width=0.8\textwidth]{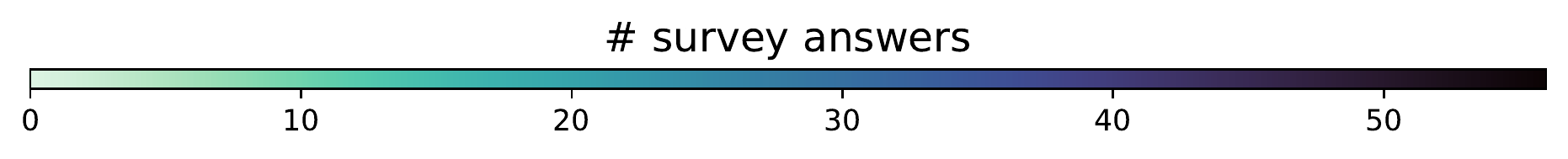}
    \label{fig:RQ1colorbar}
    \caption{Heatmaps visualizing the interpretations of the participants per study object [$S_n$].}
    \label{fig:RQ1heatmaps}
     \vspace{-.5cm}
\end{figure*}

We first look at the total number of answers for each dimension across all data sets. Secondly, we analyze the distribution of ratings based on our constructed heatmaps (see Fig.~\ref{fig:RQ1heatmaps}).

\textbf{Necessity}
Our participants did not have a clear tendency whether an antecedent is only \textit{sufficient} or also \textit{necessary} for the consequent. Among the total of 1,248 answers, 2.1\% correspond to the level ``nevertheless'', 46.9\% to ``also necessary'', and 51\% for ``only sufficient''. That means that more than half of the respondents stated that the conditional does not cover how the system is expected to work if the antecedent does not occur (i.e, the negative case is not specified).

\textbf{Temporality}
We found that time plays a major role in the interpretation of conditionals in requirements. Among the 1,248 answers, only 13\% were ``temporal ordering is irrelevant'' for the interpretation. This indicates that conditionals in requirements require temporal logics for a suitable formalization. For some study objects, the exact temporal relationship between antecedent and consequent was ambiguous. For S3, 34 participants selected  ``during'', 43 ``next state'', and 19 ``eventually''. Similarly, we observed divergent temporal interpretations for S2, S5, S7, S10, S11, and S12. In contrast, the respondents widely agreed on the temporal relationship of S1 (67 survey answers for ``next state''), S4 (84 survey answers for ``during''), S6 (73 survey answers for ``during''), S8 (67 survey answers for ``eventually'') and S9 (83 survey answers for ``eventually''). Across all study objects, 29.8\% of survey answers were given for the level ``during'', 20.1\% for ``next state'' and 37.1\% for ``eventually''. 

\textbf{Agreement}
Our heatmaps illustrate that there are only few study objects for which more than half of the respondents agreed on a 2-tuple (see Fig.~\ref{fig:RQ1heatmaps}). This trend is evident across all data sets. The presence or absence of domain knowledge does not seem to have an impact on a consistent interpretation. The greatest agreement was achieved in the case of S1 (48 survey answers for $\langle$necessary, next state$\rangle$), S6 (49 survey answers for $\langle$necessary, during$\rangle$), S8 (53 survey answers for $\langle$sufficient, eventually$\rangle$) and S9 (56 survey answers for $\langle$sufficient, eventually$\rangle$). However, for the majority of study objects, there was no clear agreement on a specific 2-tuple. For S5, two 2-tuples were selected equally often, and for S10, the two most frequent 2-tuples differed by only two survey answers. 

\textbf{Generally Valid Formalization?}
Mapping the most frequent 2-tuples in the heatmaps to our constructed formalization matrix reveals that all study objects can not be formalized in the same way. The most frequent 2-tuples for each study object yield the following six patterns:

\begin{compactenum}
  \item $\langle$necessary, next state$\rangle$: S1, S3
  \item $\langle$necessary, irrelevant$\rangle$: S2
  \item $\langle$necessary, during$\rangle$: S6, S10, S11
  \item $\langle$necessary, eventually$\rangle$: (S5)
  \item$\langle$sufficient, eventually$\rangle$: (S5), S7, S8, S9
  \item $\langle$sufficient, during$\rangle$: S4, S12
\end{compactenum}

\begin{figure*}
    \centering
    \scriptsize
    \begin{tabular}{c|c|c}
        & \textbf{Temporality} & \textbf{Necessity}  \\ \hline
    
        \rotatebox[origin=c]{90}{if} &
        \raisebox{-.54\height}{\includegraphics[width=0.45\textwidth]{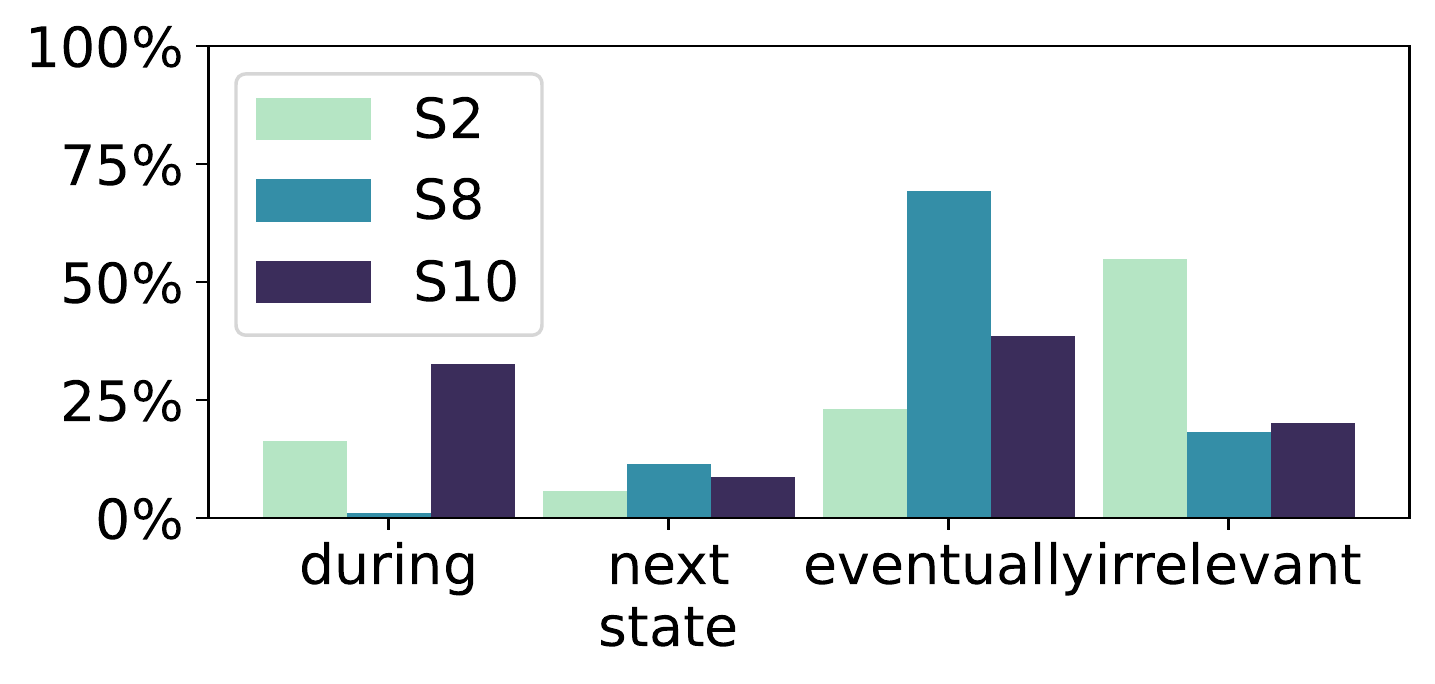}} &
        \raisebox{-.5\height}{\includegraphics[width=0.45\textwidth]{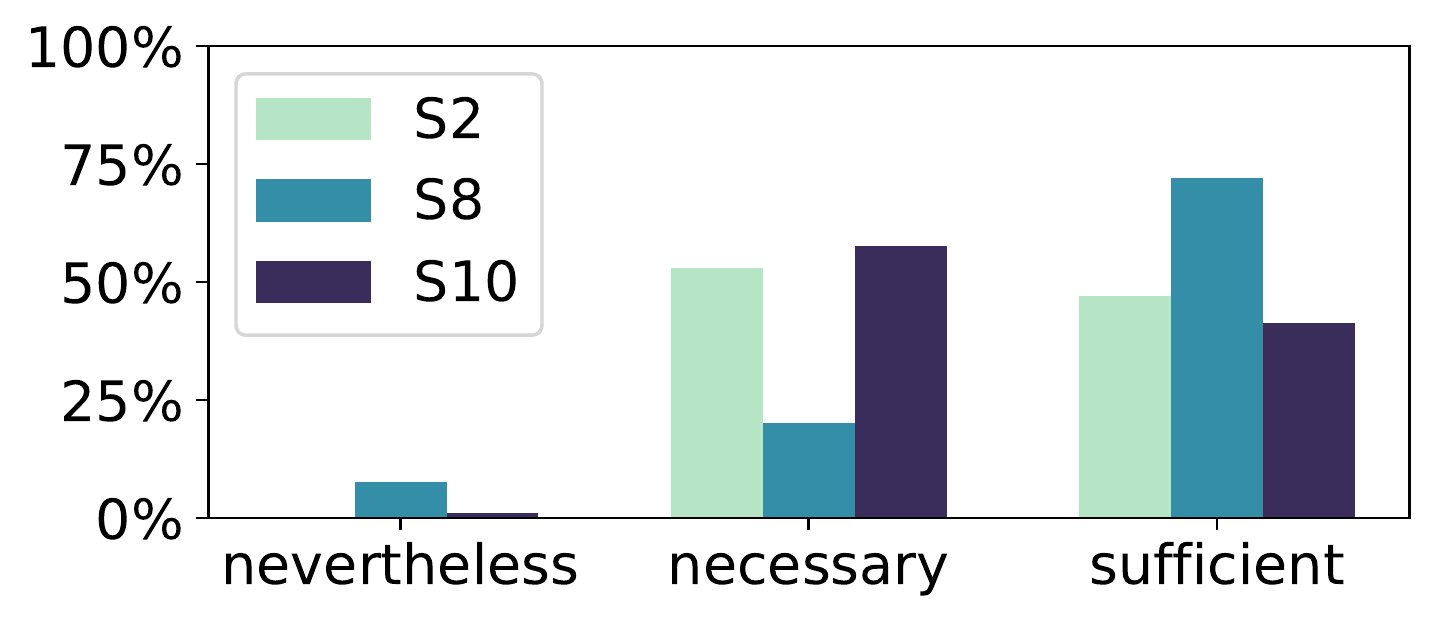}} \\\hline
        
        \rotatebox[origin=c]{90}{when} &
        \raisebox{-.54\height}{\includegraphics[width=0.45\textwidth]{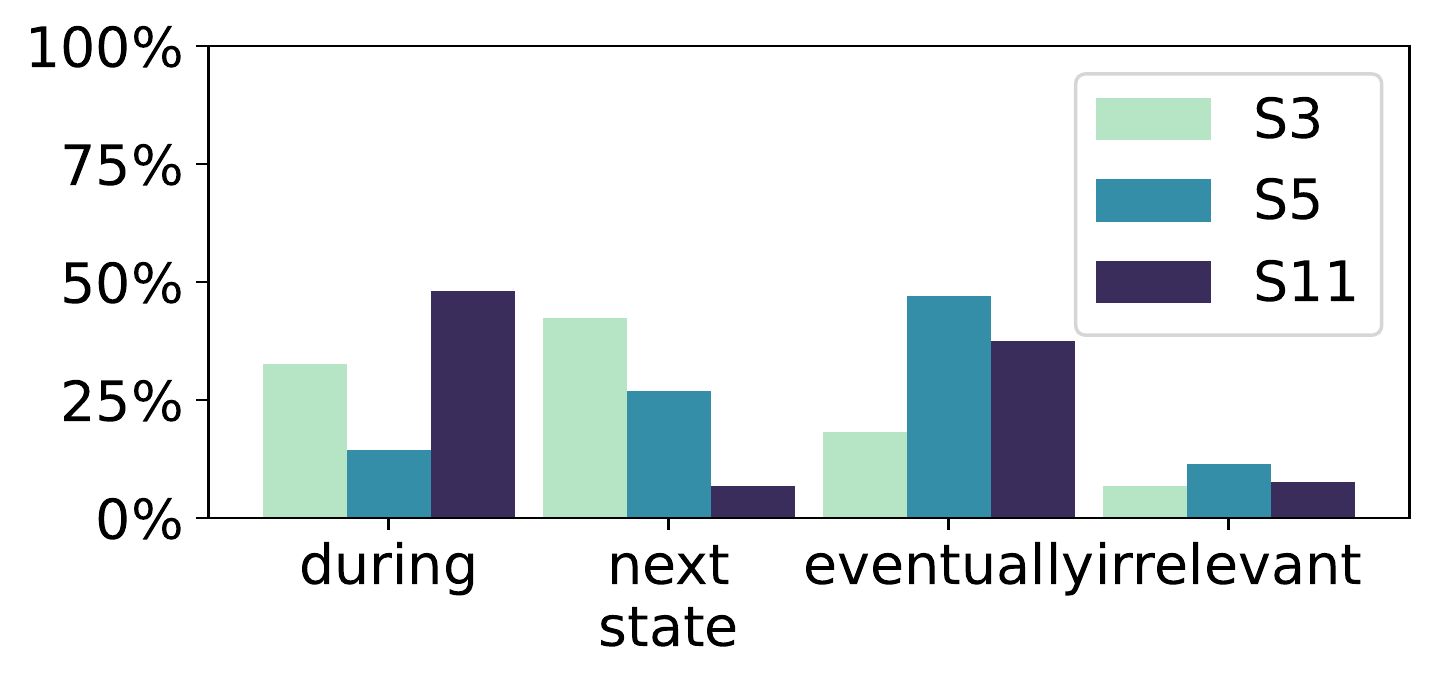}} &
        \raisebox{-.5\height}{\includegraphics[width=0.45\textwidth]{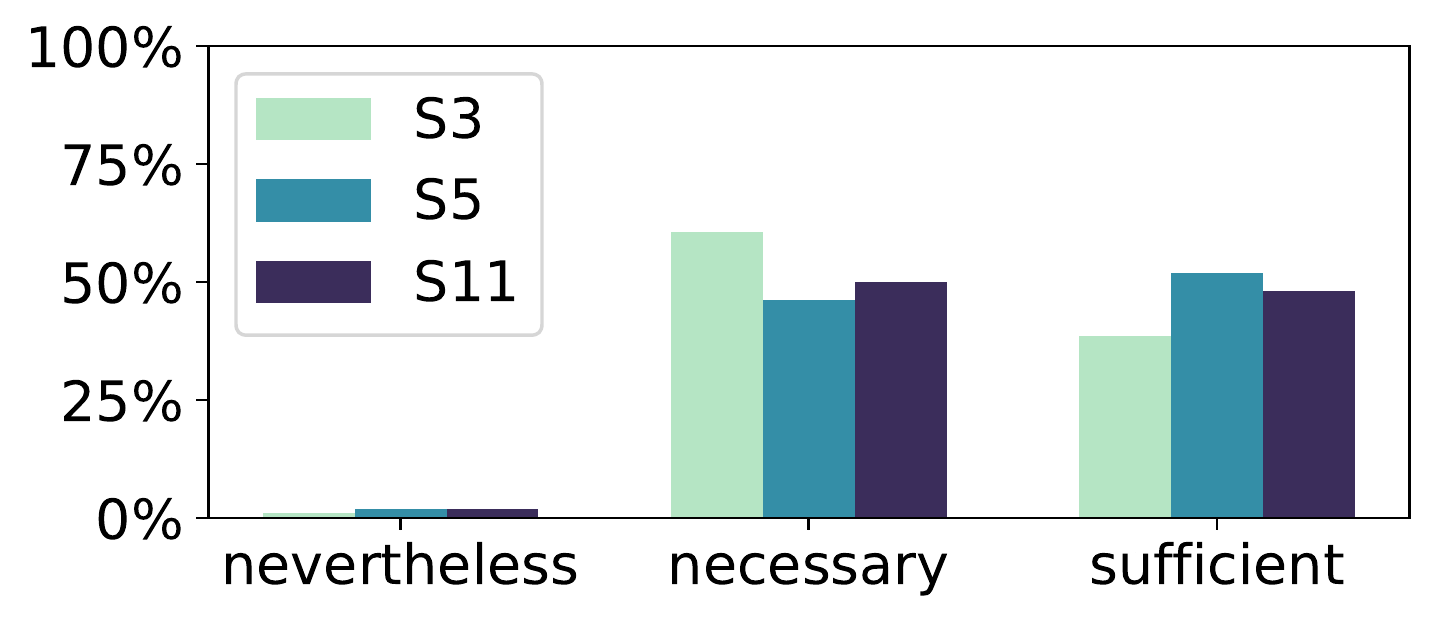}} \\\hline
        
        \rotatebox[origin=c]{90}{after} &
        \raisebox{-.54\height}{\includegraphics[width=0.45\textwidth]{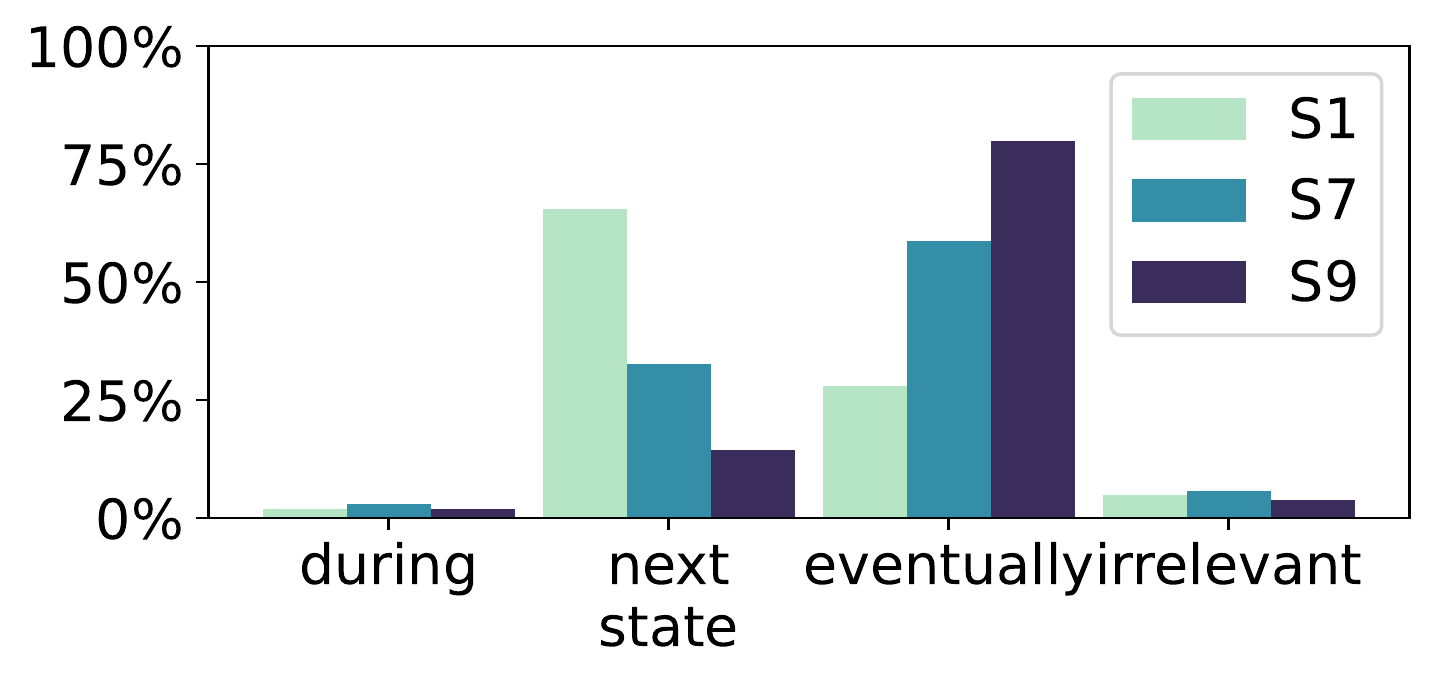}} &
        \raisebox{-.5\height}{\includegraphics[width=0.45\textwidth]{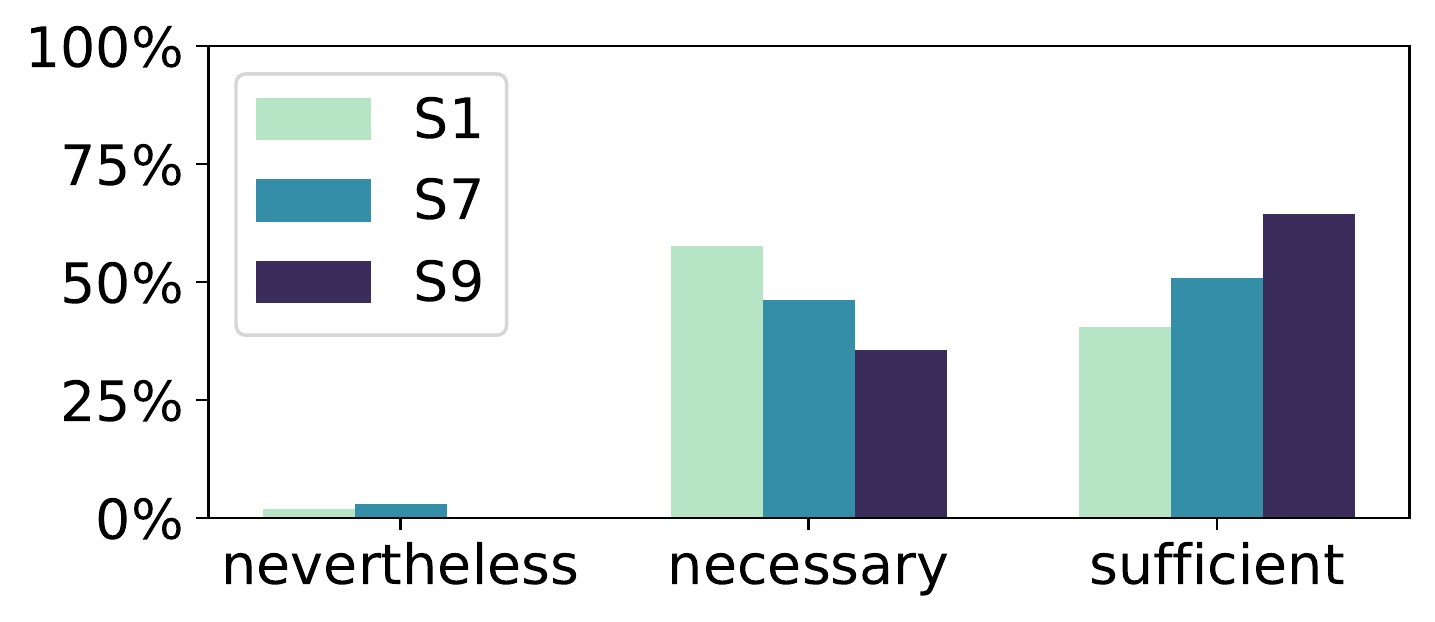}} \\\hline
        
        \rotatebox[origin=c]{90}{while} &
        \raisebox{-.54\height}{\includegraphics[width=0.45\textwidth]{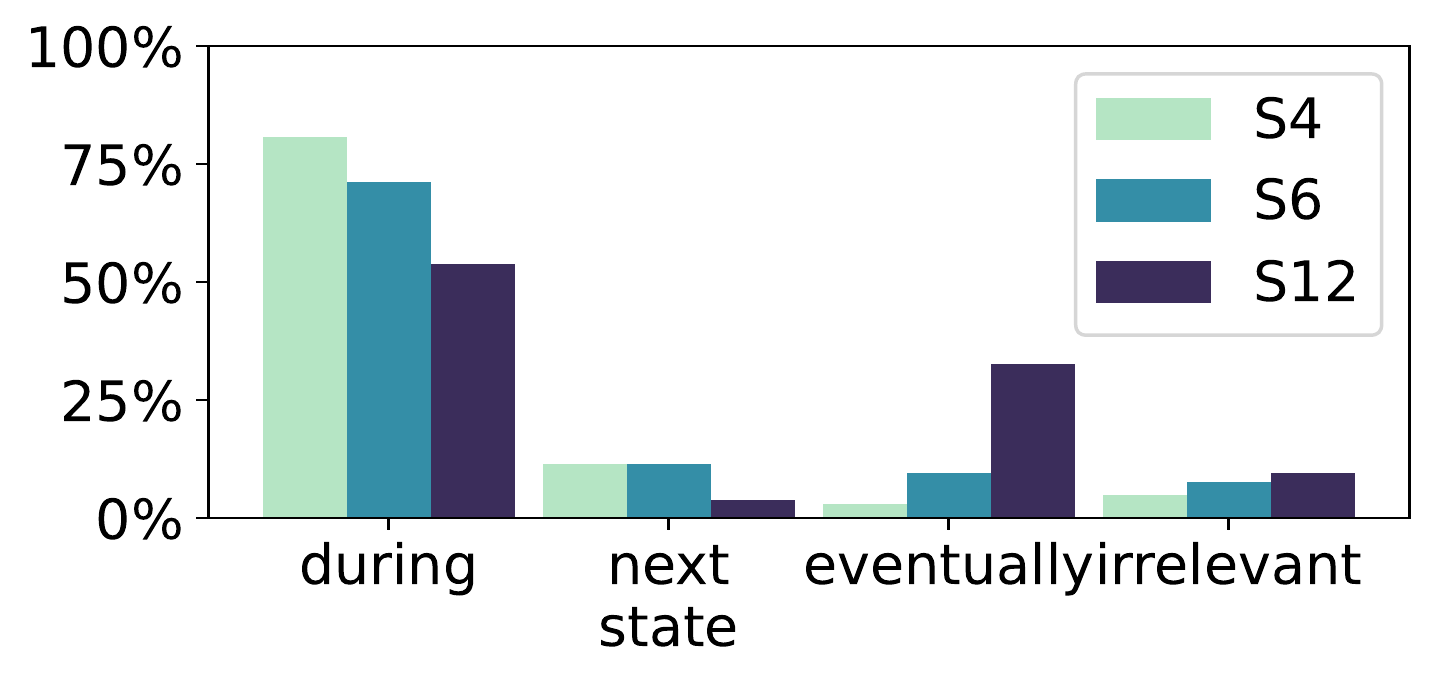}} &
        \raisebox{-.5\height}{\includegraphics[width=0.45\textwidth]{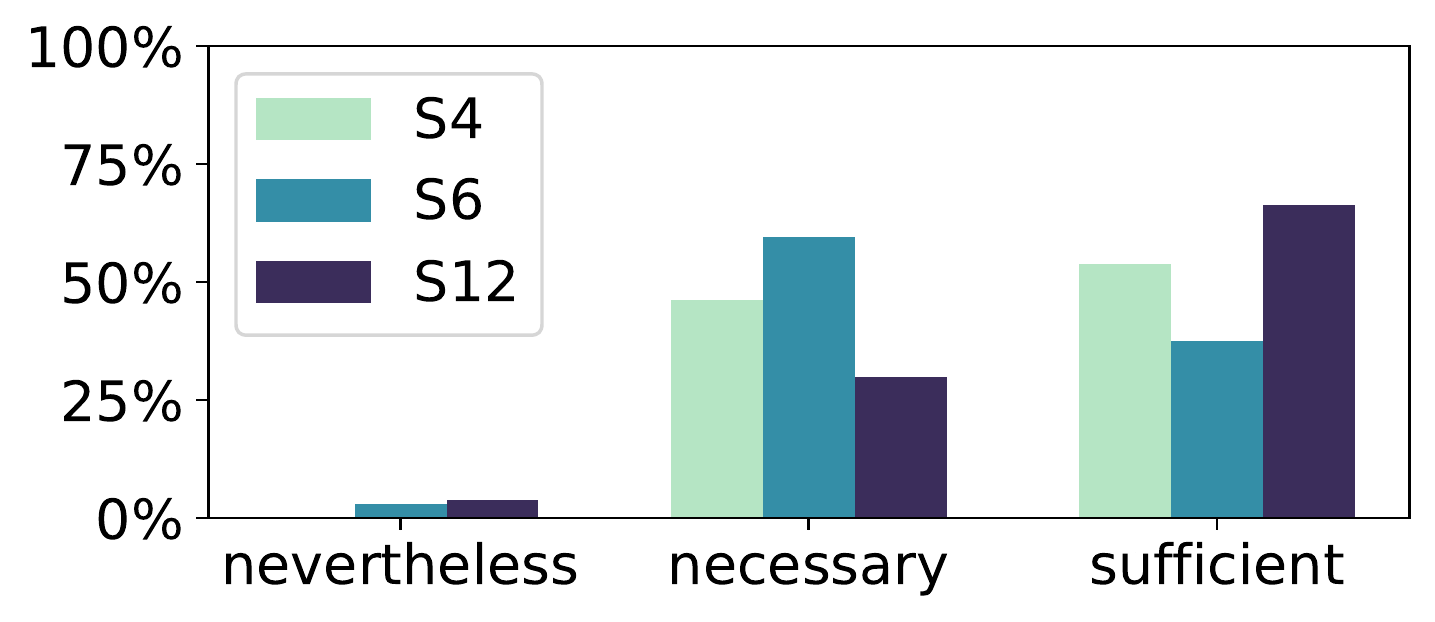}} \\
        
    \end{tabular}
    \caption{Distribution of survey answers on the different variable levels for each set of study objects with the same cue phrase (e.g., S2, S8 and S10 include ``if''). }
    \label{fig:RQ3histograms}
     \vspace{-.5cm}
\end{figure*}

One sees immediately that it is not possible to derive a formalization for conditionals in general. Especially the temporal interpretations differed between the conditionals and the used cue phrases (see Fig.~\ref{fig:RQ3histograms}). However, it can be concluded that, except for S2, the interpretations of all study objects can be represented by LTL. 
\vspace{-.5cm}

\subsection*{RQ 2: Which factors influence the logical interpretation of conditional clauses in requirements?}
\begin{table}
\scriptsize
\centering
\setlength{\extrarowheight}{0pt}
\addtolength{\extrarowheight}{\aboverulesep}
\addtolength{\extrarowheight}{\belowrulesep}
\setlength{\aboverulesep}{0pt}
\setlength{\belowrulesep}{0pt}
\caption{Relationships between factors and interpretation.}
\label{tab:testresults}
\begin{tabular}{lrrrrr} 
\toprule
                                                          & \multicolumn{3}{c}{\textbf{Test Statistics }} & \multicolumn{2}{c}{\textbf{Measures}}  \\
\cmidrule(lr){2-4}\cmidrule(lr){5-6}
\textbf{Tested Relationship}                              & $\chi^2$ & \textbf{df} & \textbf{p-value}        & $\phi$ & $\Theta$                          \\
\midrule
\rowcolor[rgb]{0.89,0.89,0.89} Experience and Necessity   & 2.384       & 6           & 0.881             & - & -                  \\
\rowcolor[rgb]{0.89,0.89,0.89} \textbf{Experience and Temporality} & 31.523      & 9           & 
0.001            & -            & 0.089~                                  \\
Interaction and Necessity                                 & 11.005           & 8           & 2.201                & -          & -                                       \\
\textbf{Interaction~and Temporality}                               & 36.991           & 12           & < 0.001                 & 0.510            & -                                      \\
\rowcolor[rgb]{0.89,0.89,0.89} \textbf{Domain and Necessity}                                      & 22.310            & 4          & < 0.001                  &   0.134        & -                                      \\
\rowcolor[rgb]{0.89,0.89,0.89} \textbf{Domain and Temporality}                                    & 138.128           & 6           & < 0.001                 & 0.333            & -                                       \\
\bottomrule
\end{tabular}
 \vspace{-.5cm}
\end{table}

This section reports the results of our chi-square tests (see Tab.~\ref{tab:testresults}). In our contingency tables, no more than 20\% of the expected counts are $<5$. Hence, we satisfy the assumption of enough observations per category for the chi-square test~\cite{Yates99}. In the following, we explain the relationships where the chi-square test indicated a dependency between the logical interpretation and a factor. 

\textbf{The logical interpretation regarding Temporality depends on RE Experience} 
In the group with less than 1 year of experience, there is a tendency to perceive the temporal relationship between the events as ``during'' (36.4\%). In the group of participants with 4--10 years of experience, most of the respondents rated the temporal relationship as ``eventually'' (41.3\%). The $\chi^2$ test reveals that the distribution of ratings differs between the experience levels. The calculated $\Theta$ value indicates that the strength of the relationship is low.

\textbf{The logical interpretation regarding Temporality is dependent on how a practitioner interacts with requirements} 
Our contingency table reveals that the distribution of ratings differs between the interaction levels. Practitioners who implement requirements fluctuate mainly between ``during'' and ``eventually'', while they rarely selected the other two Temporality levels. A different pattern emerges for practitioners who maintain and verify requirements. Across all study objects, they choose the levels ``during'', ``next state'' and ``eventually'' equally often. A $\chi^2$ test indicates a dependency between both variables. The calculated $\phi$ value indicates that the strength of the relationship is high.

\textbf{The logical interpretation regarding Necessity is dependent on domain knowledge} 
The disagreement about whether an antecedent is only \textit{sufficient} or also \textit{necessary} holds regardless of domain knowledge. However, the trend differs between the data sets with respect to the Necessity levels. In the case of DS1 (domain knowledge assumed), more answers were given for ``also necessary'' (54.3\%) than for ``only sufficient'' (45\%). In contrast, more ratings were given for ``only sufficient'' in the case of DS2 (53.1\%) and DS3 (55\%). The slight difference in the distribution of the ratings regarding Necessity is supported by the $\chi^2$ test. However, the strength of the relationships is low. 

\textbf{The logical interpretation regarding Temporality is dependent on domain knowledge}
Our contingency table shows that the distribution of ratings regarding Temporality differs between the data sets. In the case of DS1, ratings were mainly given for ``during'' (32.9\%) and ``next state'' (31.3\%). In the case of the unknown domain (DS2), ratings were mainly assigned to ``eventually'' (46.2\%), while only 20.7\% were given to ``next state'' and 22.4\% to ``during''. In DS3, where no domain knowledge is necessary for the understanding of the conditionals, most ratings were given to ``during'' (34.1\%) and ``eventually'' (47.1\%). A $\chi^2$ test shows that there is a statistically significant dependency between both variables. According to the calculated $\phi$ value, the strength of the relationship is medium.

\subsection*{RQ 3: Which (if any) cue phrases promote (un)ambiguous interpretation?}
The histograms in Fig.~\ref{fig:RQ3histograms} show that the logical interpretation regarding Temporality depends on the cue phrase used to express a conditional. For study objects containing ``while'' (S4, S6 and S12), the respondents largely agreed that the consequent occurs simultaneously with the antecedent. 
In contrast, almost no respondent associated simultaneous events in the study objects with the cue phrase ``after''. Instead, the respondents vacillated between the temporal levels ``next state'' and ``eventually''. 
The largest disagreement, though, was found in the interpretations of the conditionals ``if'' or ``when''. Especially in the case of ``when'', there was no clear agreement across S3, S5 and S11 on whether antecedent and consequent are in a  ``during'', ``next state'' or ``eventually'' temporal relationship. Regarding Necessity, we observe that the practitioners, irrespective of the used cue phrase, disagree whether the antecedent is only \textit{sufficient} or also \textit{necessary} for the consequent. We found one outlier in our histograms (S8), where an 80\% agreement for the level ``sufficient'' could be achieved. For the remaining study objects, however, there is a balanced number of survey answers for both levels. 

\section{Threats to Validity}
\textbf{Internal Validity}
The respondents may have misunderstood the questions resulting in poor quality or invalid answers. To minimize this threat, we followed the guidelines by Dillmann~\cite{Dillman14} in the creation of the questionnaire. In addition, we conducted a pilot phase to validate the questionnaire internally through discussions in the research team and externally through pilot survey runs. Selection bias is another threat. Although we have started with personal contacts to find participants, the sampling process has been extended by indirect contacts. As a result, selection bias has been reduced. Another possible threat is the selection of dimensions by which we formalize conditionals. The two dimensions, Temporality and Necessity, have been selected after extensive literature research and discussion among the authors. However, the completeness of dimensions can neither be proven nor rebutted. One threat that we were unable to control was the distribution of native speakers. Although one could argue that non-native speakers reading and writing English requirements are the standard case for most projects, and therefore, their interpretation is meaningful nevertheless, future research should validate the findings also with a dedicated group of native speakers. 
Another threat arises from our assumption that each participant has the necessary domain knowledge in case of DS1 but lacks it in case of DS2. To mitigate this threat, we analyzed the feedback received during our pilot study. In the case of DS1, almost no questions were raised, whereas in the case of DS2, many pilot users lacked knowledge about the described system behavior. This indicates that the respondents may have the necessary domain knowledge to interpret the conditionals described in DS1. Furthermore, the conditionals in DS1 are derived from the data set used in the tool competition at the NLP4RE workshop, which is claimed to be interpretable without specific domain knowledge.

\textbf{External Validity}
As in every survey, the limited sample size and sampling strategy do not provide the statistical basis to generalize the results of the study. However, we tried to involve RE practitioners working in different roles at companies from different domains to obtain a comprehensive picture of how conditionals are logically interpreted. We argue that our survey sample of 104 RE practitioners, who work in 22 different domains and of which a third have more than 10 years of experience in RE is sufficient for a first insight into the logical interpretation of conditionals. 

\textbf{Construct Validity}
The questionnaire might not sufficiently cover our research questions limiting the availability of data that provides suitable answers to the research questions. To minimize this threat, we constructed a formalization matrix and designed our questionnaire according to the dimensions of the matrix to establish a distinct mapping between interpretation and suitable formalization. 

 
\section{Concluding Discussion and Outlook}
Conditionals are common to specify desired system behavior. In this paper, we show that conditionals are interpreted ambiguously by RE practitioners. In particular, there is disagreement (1) about whether an antecedent is only \textit{sufficient} or also \textit{necessary} for a consequent, and (2) about the temporal occurrence of antecedent and consequent when different cue phrases (such as ``when'' or ``if'') are used. Thus, a generic formalization of conditionals will inevitably fail at least some practitioner's interpretation. We see two immediate implications in practice:

\textbf{(1) Implications for automatic methods} Especially (if not limited) for automated test case generation, it is vital to understand which behavior is desired if the antecedent does not occur. The evidence presented in this paper refutes the prevailing assumption (cf.~\cite{mavin09,fischbachRE}) that antecedents can always be treated as \textit{necessary} conditions.
Hence, we propose that future methods should display the automatically generated positive and negative test cases to practitioners and explicitly verify: ``Is the negative case of your conditional also valid?''. This will foster the discussion within project teams about the expected system behavior and enables to resolve misunderstandings at an early stage.

\textbf{(2) Implications for requirements authors} It should be incorporated into RE writing guidelines that it does matter which cue phrase is used for the formulation of a conditional. ``While'' is interpreted consistently, but ``if'' and ``when'' cause misunderstandings about the temporal interpretation of antecedent and consequent. This poses a problem especially in the implementation of requirements and eventually leads to discrepancies between actual and expected system behavior. Project teams should therefore agree early on how they want to interpret the different cue phrases to avoid ambiguities. Additionally, our findings provide empirical evidence for the claim by Berry~et~al.~\cite{Berry03} and Rosadini~et~al.~\cite{Rosadini17} that requirements authors should always specify the negative case (e.g., by using an else-statement) to prevent confusion about the necessity of antecedents.

Our observations open an avenue for further investigations. Among them, we believe that it is interesting to explore how other cue phrases (e.g. once, because, as soon as) are interpreted logically. It would furthermore be interesting to compare our results with the logical interpretation of requirements written in other languages. 

%
%
\bibliographystyle{splncs04} 
\bibliography{references}
\end{document}

%% file: commands.tex
\newcommand{\cornItIII}{$\square ( F \Rightarrow \lozenge G)$}

\newcommand{\cornItV}{$\square ( F \Rightarrow G)$}

\newcommand{\cornIItIV}{$\square((\neg F \Rightarrow \neg (\lozenge G)) \land (F \Rightarrow \ocircle G))$}

\newcommand{\cornIItV}{$\square((\neg F \Rightarrow \neg (\lozenge G)) \land (F \Rightarrow G))$}

\newcommand{\cornIItVI}{$F \Leftrightarrow G$}